\documentclass[12pt]{article}
\usepackage{geometry}
\usepackage{a4}
\usepackage{graphicx}
\usepackage{epsf}
\usepackage{amsmath}
\usepackage{amssymb}
\usepackage[T1]{fontenc}
\usepackage[utf8]{inputenc}
\usepackage{tabularx,ragged2e,booktabs,caption}
\newcolumntype{C}[1]{>{\Centering}m{#1}}

\usepackage{cite}
\newcommand{\be}{\begin{equation}}
\newcommand{\ee}{\end{equation}}

\newcommand{\Rmnum}[1]{\expandafter\@slowromancap\romannumeral #1@}
\newcommand{\bea}{\begin{eqnarray}}
\newcommand{\eea}{\end{eqnarray}}

\begin{document}
\def\A{{\mathbb{A}}}
\def\B{{\mathbb{B}}}
\def\C{{\mathbb{C}}}
\def\R{{\mathbb{R}}}
\def\s{{\mathbb{S}}}
\def\T{{\mathbb{T}}}
\def\Z{{\mathbb{Z}}}
\def\W{{\mathbb{W}}}
\begin{titlepage}
\title{Generalized Superconductors and Holographic Optics - II}
\author{}
\date{
Subhash Mahapatra
\thanks{\noindent E-mail:~ subhmaha@iitk.ac.in}
\vskip0.4cm
{\sl Department of Physics, \\
Indian Institute of Technology,\\
Kanpur 208016, India. }}
\maketitle
\abstract{Using linear response theory, we analyze optical response properties of
generalized holographic superconductors, in AdS-Schwarzschild and single
R-charged
black hole backgrounds in four dimensions. By introducing momentum
dependent vector mode perturbations, the response functions for these
systems
are studied numerically, including the effects of backreaction. This
complements and completes the probe limit analysis for these backgrounds
initiated in our previous work ({\tt arXiv : 1305.6273}).
Our numerical analysis indicates a negative Depine-Lakhtakia index for
both the backgrounds studied, at low enough frequencies. The dependence of
the response functions
on the backreaction parameter and the model parameters are established and
analyzed with respect to similar backgrounds in five dimensions.}
\end{titlepage}

\section{Introduction}

The AdS/CFT correspondence provides an astonishing duality between classical gravity in $(d+1)$ dimensional AdS spacetime and a
conformal field theory (CFT) living at the boundary of the $(d+1)$-dimensional AdS space \cite{Maldacena}\cite{Klebanov}\cite{Witten}.
For the last few years, this has been recognized as a powerful tool to study strongly coupled field theories, and there is hope that AdS/CFT can be an effective tool
in understanding physical phenomena in condensed matter systems. The primary reason for the usefulness of the AdS/CFT correspondence lies in the fact
that it provides a unique approach to address some questions in strongly coupled field theories which otherwise would be intractable. Indeed by now, several exciting directions
of research from AdS/CFT correspondence have emerged and there are indications that for strongly coupled condensed matter systems, realistic predictions might be possible.
For example, a few areas in which the AdS/CFT duality find applications are non-Fermi liquids, quantum quenches, superconductivity etc.
The purpose of the present paper is to study optical response properties in strongly coupled systems, that in many cases mimic actual experimental configurations. In particular,
we will address this issue in the context of generalized holographic superconductors.

It is by now well known that the Abelian Higgs Lagrangian in the background of AdS black holes provides the holographic description of the superconductivity \cite{Gubser}\cite{Hartnoll}.
It was shown by Gubser \cite{Gubser} that the $U(1)$ gauge symmetry in the AdS black hole backgrounds can be broken spontaneously  below a
critical temperature and that the AdS black holes support scalar hair below this temperature. The main reason for the scalar field instability is the minimal
coupling between the gauge field and the scalar field, which adds a negative term to the mass of the scalar field and makes the effective mass of the scalar field
sufficiently negative near the black hole horizon. In the AdS/CFT language, the spontaneous breaking of $U(1)$ gauge symmetry in the bulk corresponds to
the global $U(1)$ symmetry breaking at the boundary. This in turn implies a nonzero vacuum expectation value of the charged scalar operator, which is
dual to the scalar field in the bulk, and a phase transition to the superconducting phase at the boundary\cite{Hartnoll}. The boundary systems obtained in this
way from AdS/CFT correspondence possess all the main characteristic properties of superconductivity like infinite DC conductivity, energy gap, Meissner like effects
etc (see e.g. \cite{Hartnoll1}). An important generalization in the theory of holographic superconductors was done in \cite{Franco} where the authors
introduced a non-minimal interaction between the gauge and the scalar field in a gauge invariant way and broke the $U(1)$ symmetry spontaneously by
a Stuckelberg mechanism. The superconductors constructed in this way are called the generalized holographic superconductors. Importantly, these
generalized superconductors exhibit richer phase structure than the minimally coupled superconductors, in particular one can now control the order of phase
transition and the critical exponents. This is important from a phenomenological as well as experimental point of view since there are superconductors which show
first order phase transitions\cite{Bianchi}. In real superconductors, such phase transitions occur in the presence of external magnetic fields which is not the case
that we consider. Nevertheless, we expect that our strong coupling description should provide important predictions for real systems, possibly in future experiments.

In this paper, we will primarily focus on an exotic property exhibited by some materials - negative refraction.
This is an area of active research in optics, that started soon after the discovery of a new class of artificial media commonly called ``metamaterials'' which exhibit an
unusual property of wave propagation inside the medium\cite{Smith}\cite{Pendry}. Namely, here the energy flow (as dictated by the Poynting vector) propagates in the
direction opposite to the wave vector or phase velocity. Without frequency dispersion, this happens when the medium shows simultaneous negative
values of the electric permittivity $\varepsilon$ and the magnetic permeability $\mu$. Since wave propagation inside the medium is generally characterised
by the refractive index, $n=\pm\sqrt{\varepsilon \mu}$, negative refraction implies choosing the negative sign in $n$. However, taking frequency
dispersion into account, which makes $\varepsilon$, $\mu$ and $n$ complex, the refraction can also become negative if $Re(\varepsilon)$ and $Re(\mu)$
are not simultaneously negative. The relevant parameter used to establish the phenomenon of negative refraction in the medium is called the Depine-Lakhtakia (DL) index
$n_{DL}$, with negativity of the DL index indicating that the direction of energy flow is opposite to the direction of phase velocity in the medium \cite{Depine}.
We will elaborate on the essential conditions for negative refraction more in section $2$.  This peculiar behaviour of wave propagation lead to the modification
of many laws of physics such as Snell's law, Doppler effect etc. More details on the properties of the medium with negative refraction along with historical background can
be found in \cite{Agranovich}.

Inspired by the properties of metamaterials, the work of \cite{Policastro} initiated its study in the context of the gauge-gravity duality. These authors showed that the
boundary theory corresponding to an Einstein-Maxwell bulk theory at finite chemical potential and temperature in the five dimensional AdS black hole background
have $n_{DL}<0$ and shows negative refraction in the low frequency regime. It was then subsequently investigated in RN-AdS black hole in four
dimensions \cite{Ge}, R-charged black holes in different dimensions\cite{Prabwal} and $D7$ flavor brane systems\cite{Tarrio}. Identical results of negative refraction in
low frequency were found in all cases. There is a caveat that we have to keep in mind, namely that there is no dynamical photon in the boundary theory. That is, the
strongly coupled systems that we consider are assumed to have a weak coupling with a dynamical photon, that is treated perturbatively.

The optical properties of holographic superconductors in Einstein $+$ Maxwell $+$ scalar field theory have been studied for a few cases.
In part, this is motivated by experimental results in some superconductors which show negative refraction\cite{Anlage}. In \cite{Gao}, it was shown that holographic
superconductors in $(3+1)$ dimension in the probe limit always have $n_{DL}>0$ and do not exhibit negative refraction at any frequency.
However, by considering the backreaction of the scalar and the gauge field on the AdS-Schwarzschild black hole, it was explicitly shown in \cite{Amariti}
that holographic superconductors do exhibit negative refraction in low frequency regions. The reason for the emergence of negative refraction with backreaction
is the  appearance of an extra diffusive pole in the current current correlators due to metric fluctuations. Subsequently, the optical properties
of $(2+1)$ dimensional holographic superconductors in R-charged and AdS-Schwarzschild black holes were studied in \cite{Subhash},\cite{Dey}, where it was
found that $n_{DL}$ can be negative even in the probe limit. Later on, it was pointed out in \cite{Dey} that negative $n_{DL}$ may not be a good
criteria for negative refraction when $Im(\mu)$ is negative, which generally occurs in the probe limit. Here, it was shown that for
some special conditions, negative $n_{DL}$ can occur along with positive $Im(\mu)$ in the probe limit. Now it becomes clear that in terms of
negative refraction, holographic superconductors in $(3+1)$ and $(2+1)$ dimensions show different behavior in the probe limit. Therefore,
it is important to investigate the refractive index and the other response functions of $(2+1)$ dimensional holographic superconductors including the
effects of backreaction, in order to draw any definitive conclusion.

In this work, we study the optical properties of generalized holographic superconductors in $(2+1)$ dimensions including effects of backreaction.
We consider two different four dimensional bulk backgrounds, namely the AdS-Schwarzschild and the single R-charged black hole backgrounds in four dimensions.
The main results of this paper are summarized below. \\
\noindent
$\bullet$ We find that as we turn on the backreaction, the metric fluctuations introduce an additional pole in the current-current
correlator which makes the imaginary part of the magnetic permeability positive in the small frequency region, contrary to the probe limit. \\
$\bullet$ By calculating $n_{DL}$, we find that negative refraction generally occurs with backreaction  at sufficiently small frequencies for both backgrounds.
This is again contrary to the result in the probe limit. For the AdS-Schwarzschild black hole background, our findings suggest that holographic
superconductors with higher backreaction support negative refraction for a wider frequency range, which is qualitatively similar to the results of $3+1$ dimensional
holographic superconductors.  However curiously, the same is not true in the R-charged black hole background,
we find that the frequency range where $n_{DL}<0$ is nearly independent of the backreaction parameter. Our results also strengthen the general claim made in \cite{Amaritihydro}
that the hydrodynamic systems which have gravity duals normally exhibit negative refraction below certain cutoff frequency.\\
$\bullet$ In both the backgrounds considered, we find negative refraction for all temperatures. By comparing with the results for the normal phase,
we find qualitatively similar nature of the response functions for the normal and the superconducting phase at a particular temperature, although the
magnitudes of the response functions in these two phases are different.  \\
$\bullet$ It is shown that the propagation to the dissipation ratio is always negative in the frequency range where negative refraction occurs.
Further, as is normally the case in metamaterials, we find large dissipation in the system, however higher backreaction seems to enhance the propagation.\\
$\bullet$ We establish the dependence of response functions on the model parameters and find that the transition from positive to negative refraction
with frequency is almost independent of $\xi$. However, the higher values of $\xi$ increases the dissipation in the system. This could be phenomenologically
important as it opens up the possibility of tuning the magnitude of the dissipation by introducing additional model parameters which may provide some insight
 for improving the propagation in the metamaterials. This can have many physical applications.

This paper is organised as follow. In section $2$, we review the basics of linear response theory and state the necessary formulas for response functions.
This will set the basic notations for rest of the paper. In section $3$, we first describe the holographic set up under consideration and then proceed to
show the superconducting nature of the boundary theory. We will work in 4D AdS-Schwarzschild background with full backreaction .
The general procedure to calculate response functions via AdS/CFT correspondence and the numerical results are outlined in section $4$.
We will adopt the method used in \cite{Amariti}. In section $5$, we will state the results of optical properties in 4D single R-charged black hole background.
We end with our conclusions and some discussions along with future directions in section $6$.\footnote{We are very grateful to A. Amariti and D. Forcella
for generously sharing their Mathematica code that computes response functions in $3+1$ dimensional
holographic superconductors. The computations in this paper have been performed using a somewhat different Mathematica routine that is inspired from this,
and is available from the author on request.}

\section{Linear response theory and negative refraction}

In this section we will briefly discuss some basic aspects of linear response theory and establish the formulas which we will require to calculate the
optical properties of the boundary system.

It is well known that the electromagnetic response of a continuous medium is completely characterized by its electric permittivity ($\varepsilon$)
and magnetic permeability ($\mu$). There are usually two different approaches in dealing with wave propagation in continuous media \cite{Landau}.
The first approach, which is generally called as the $\varepsilon-\mu$ approach, involves the set of fields $E$, $D$, $B$ and $H$ which are related to
each other by the Maxwell's equations and the matter equations. The latter, which also describe the response of the medium, are
\begin{eqnarray}
D_i(\omega)=\varepsilon_{ij}(\omega)E_{j}, \ \ \ B_i(\omega)=\mu_{ij}(\omega)H_{j}
\end{eqnarray}
Here $\varepsilon_{ij}$ and $\mu_{ij}$ encode the information about the response of the medium and in an isotropic medium they reduce
to scalar functions ($\varepsilon_{ij}=\varepsilon \delta_{ij}$ and $\mu_{ij}=\mu \delta_{ij}$). One observation from these relations is that,
in this approach the response functions only depend on the frequency $\omega$ and they are in general complex quantities.

The second, more general approach is called the spatial dispersion approach, where we explicitly take the  dependence of the wave vector $k$ in the
response functions. This involves the set of fields $E$, $D$, $H = B$, satisfying the Maxwell's equations but the matter equations are now
\begin{eqnarray}
D_i(\omega)=\varepsilon_{ij}(\omega,k)E_{j}
\label{spatialresps}
\end{eqnarray}
In this case, the optical response of the medium is completely characterized by the generalized dielectric tensor $\varepsilon_{ij}(\omega,k)$,
a function of the wave vector $k$. As is clear from the definition, $\varepsilon_{ij}(\omega,k)$ contains both the electric as well as the magnetic response.
Now, decomposing $\varepsilon_{ij}$ into the transverse part $\varepsilon_T$ and the longitudinal part $\varepsilon_L$
\cite{Agranovich} \footnote{We can do this for the isotropic medium which is the case under consideration.}
\begin{eqnarray}
\varepsilon_{ij}(\omega,k)=\varepsilon_{T}(\omega)\biggl(\delta_{ij}-\frac{k_i k_j}{k^2}\biggr) +\varepsilon_{L}\biggr( \frac{k_i k_j}{k^2}\biggr)
\label{transres}
\end{eqnarray}
and using Maxwell's equations \footnote{The Maxwell's equations imply $D_i=\frac{k^2}{\omega^2}\biggl(E_i-\frac{k_i(k.E)}{k^2}\biggr)$.}
along with eq. (\ref{spatialresps}), one can find the dispersion relation for the transverse and longitudinal permittivity as
\begin{eqnarray}
\varepsilon_{T}(\omega,k)=\frac{k^2}{\omega^2}=n^2, \ \ \  \varepsilon_{L}(\omega,k)=0
\label{transres1}
\end{eqnarray}
The complex quantity $n$ is the refractive index of the medium, with its imaginary part giving the information about the dissipation and
the real part encoding information about the propagation of the electromagnetic waves inside the medium.

In order to show that the two approaches mentioned above are equivalent (at least in some limit), we expand the transverse part of the dielectric
tensor for small $k$ as
\begin{eqnarray}
\varepsilon_{T}(\omega,k)=\varepsilon(\omega)+\frac{k^2}{\omega^2}\biggl(1-\frac{1}{\mu(\omega)}\biggr)+O(k^4)
\label{transres2}
\end{eqnarray}
Now with this expansion of $\varepsilon_{T}(\omega,k)$ and using eq. (\ref{transres1}), we obtain the relation $n^2=\varepsilon \mu$,
which is the usual relation for refractive index in the $\varepsilon-\mu$ approach. This indicates that these two approaches are equivalent at
least upto $k^2$ term in the expansion. However, there are a few advantages offered by spatial dispersion approach, the full details of which are
beyond the scope of this paper, and the reader is referred to \cite{Agranovich}. Eq. (\ref{transres2}) is generally taken as the definition of
$\mu(\omega)$ in the spatial dispersion approach and is called the effective magnetic permeability. It is effective in the sense that it contains
information about the magnetic as well as the electric response of the medium. We will keep this in mind when discussing the results in section $4$.

Different equivalent indicators of negative refraction exist in the literature. Here we will consider the Depine-Lakhtakia index $(n_{DL})$ defined as
\begin{eqnarray}
n_{DL}=|\varepsilon(\omega)|Re(\mu(\omega))+|\mu(\omega)|Re(\varepsilon(\omega))
\label{nDL}
\end{eqnarray}
It can be shown that $n_{DL} <0$ implies that the system has negative refractive index, so that the phase velocity in the medium is opposite
to the direction of energy flow. The condition $n_{DL} <0$ is derived from the requirement that the equations
$$Re(n)<0, \ \ \ \ Re(n/\mu)>0$$
should be satisfied simultaneously for negative refraction. Here, the first and second conditions give the direction of phase velocity
 and the energy propagation respectively inside the medium and as mentioned earlier, they should be of opposite signs for negative refraction ($n_{DL}$ >0 implies
usual positive refractive index). We should also point out here that the derivation of $n_{DL}$ in eq.(\ref{nDL}) is strictly based on the
assumption that $Im(\varepsilon)>0$ and $Im(\mu)>0$. For $Im(\mu)<0$, which generally occurs in the probe limit, negative $n_{DL}$ may
not be a good criterion to indicate the presence of negative refraction and more care has to be taken, as mentioned in \cite{Dey}.

For computational purposes, using linear response theory, we can write the response functions in terms of current-current correlators,
from which the response functions can be easily obtained. For an external field $A_i$, we can write the current in the linear response theory
as $J_i=C^2_{em}G_{ij}A_j$. Here, $C_{em}$ is the electromagnetic coupling constant and $G_{ij}$ is the retarded current-current correlator.
In terms of $G_{ij}$, the form of $\varepsilon_T$ can be specified as\cite{Landau}\cite{Dressel}
\begin{eqnarray}
\varepsilon_T(\omega,k)=1+\frac{4 \pi}{\omega^2}C^2_{em}G_{T}(\omega,k)
\label{transres3}
\end{eqnarray}
where $G_{T}$ is the transverse part of the correlator $G_{ij}$, which has same decomposition as in eq.(\ref{transres}).
Expanding $G_{T}(\omega,k)$ in terms of the wave vector $k$, \footnote{Here we are assuming small spatial dispersion.} we get
 \begin{eqnarray}
G_T(\omega,k)=G^{(0)}_{T}+k^2 G^{(2)}_{T}+O(k^4)
\label{GTexp}
\end{eqnarray}
Then, using eq. (\ref{transres2}), the permittivity and the effective permeability can be expressed in terms of $G^{(0)}_{T}$ and $G^{(2)}_{T}$ as
\begin{eqnarray}
\varepsilon(\omega)=1+\frac{4 \pi}{\omega^2} C^2_{em} G^{(0)}_T(\omega) &\nonumber \\
\mu(\omega)=\frac{1}{1-4\pi C^2_{em}G^{(2)}_T(\omega)}
\label{epsilonmu}
\end{eqnarray}
We will take above definition of $\varepsilon$ and $\mu$ to calculate $n_{DL}$ in eq.(\ref{nDL}). Now, the only quantities that we require in order to
calculate $\varepsilon$ and $\mu$ are the functions $G^{(0)}_T(\omega)$ and $G^{(2)}_T(\omega)$. Herein the holographic principle is invoked.
We will calculate these functions using the prescription of the AdS/CFT correspondence. The main procedure for calculating $G^{(0)}_T(\omega)$
and $G^{(2)}_T(\omega)$ through the AdS/CFT correspondence is elaborated in section $4$.

Before we discuss our holographic set up, a word about the validity of the expansion in eq.(\ref{GTexp}) is in order.
For this expansion to be valid we must have $|\frac{k^2G^{(2)}_T(\omega)}{G^{(0)}_T(\omega)}|\ll 1$. We will explicitly show in section $4$ that this is
indeed the case for the models that we consider, and that the expansion in eq.(\ref{GTexp}) is trustworthy. However, we should mention here that the
expansion may not always be valid in the probe limit. This situation will then generally indicate the breakdown of $\varepsilon-\mu$ approach.

\section{Holographic set up and superconductivity}

In this section we will set up a simple gravity dual for the generalized holographic superconductors in a four-dimensional AdS-black hole background.
We start with the Einstein-Maxwell scalar field action
\begin{eqnarray}
\textit{S}= \int \mathrm{d^{4}}x\! \sqrt{-g}\biggl[\frac{1}{2\kappa^{2}}\biggl(R+\frac{6}{L^{2}}\biggr)
-\frac{1}{4}\textit{F}_{\mu\nu}\textit{F}^{\mu\nu} -\frac{1}{2}|\textit{D}\tilde{\Psi}|^{2} -\frac{1}{2}m^{2}|\tilde{\Psi}|^{2}
\biggl] \,
\label{action}
\end{eqnarray}
Here, $\tilde{\Psi}$ is the complex scalar field with charge $q$ and mass $m$. $D_{\mu}=\partial_{\mu}-i q A_{\mu}$, $F=dA$,
$\kappa$ is related to the four dimensional Newton's constant and $L$ is the AdS length scale which is related to the negative cosmological constant.
Now rewriting the charged scalar field $\tilde{\Psi}$ as $\tilde{\Psi}= \Psi  e^{i \alpha}$, the action can be cast as
\begin{eqnarray}
\textit{S}= \int \mathrm{d^{4}}x\! \sqrt{-g}\biggl[\frac{1}{2\kappa^{2}}\biggl(R+\frac{6}{L^{2}}\biggr)-\frac{1 }{4}\textit{F}_{\mu\nu}\textit{F}^{\mu\nu} -\frac{(\partial_{\mu}\Psi)^2}{2}  -
\frac{m^2\Psi^2}{2}  &\nonumber \\ -\frac{\Psi^2(\partial\alpha-q A)^2}{2} \biggr] \,
\label{actionpsi2}
\end{eqnarray}
The gauge symmetry in this action is now given by $\alpha\rightarrow \alpha+q \lambda$ and $A_{\mu}\rightarrow A_{\mu}+\partial_{\mu}\lambda$.
There is however, now a possibility to generalise the action in eq. (\ref{actionpsi2}) in a gauge invariant way by replacing $\Psi^2$ in last line of
eq. (\ref{actionpsi2}) by a generalized function of $\Psi$. Following the procedure in \cite{Franco}\cite{Dey}, we can generalise the action as
\begin{eqnarray}
\textit{S}= \int \mathrm{d^{4}}x\! \sqrt{-g}\biggl[\frac{1}{2\kappa^{2}}\biggl(R+\frac{6}{L^{2}}\biggr)-\frac{1 }{4}\textit{F}_{\mu\nu}\textit{F}^{\mu\nu} -\frac{(\partial_{\mu}\Psi)^2}{2} -
\frac{m^2\Psi^2}{2}
&\nonumber \\ - \frac{|\textrm{G}(\Psi)|(\partial\alpha-q A)^2}{2} \biggr]
\label{actionpsi3}
\end{eqnarray}
Here $\textrm{G}(\Psi)$ is a generalized analytic functional of $\Psi$ whose form will be specified in the subsequent text.
It is clear from the form of the action in eq.(\ref{actionpsi3}) that it is still invariant under the gauge symmetry.

Now by varying the action, we find the equation of motion (EOM) for the gauge field as
\begin{eqnarray}
 \frac{1}{\sqrt{-g}}\partial_{\mu}\biggl(\sqrt{-g}\textit{F}^{\mu\nu}\biggr) +\biggl[\textrm{G}(\Psi)(\partial^{\nu}\alpha-A^{\nu})\biggr]=0
\end{eqnarray}
The EOM for the scalar field is
\begin{eqnarray}
\frac{1}{\sqrt{-g}}\partial_{\mu}\biggl(\sqrt{-g}\partial^{\mu}\Psi\biggr) - m^{2}\Psi - \frac{(\partial\alpha-A)^{2}}{2} \frac{d\textrm{G}(\Psi)}{d\Psi}=0
\end{eqnarray}
The Einstein equatoins reads
\begin{eqnarray}
\frac{1}{2\kappa^{2}}\biggl(R_{\mu\nu}-\frac{1}{2}g_{\mu\nu}R-3g_{\mu\nu}\biggr) + \frac{1}{8} g_{\mu\nu} \textit{F}^{2}-\frac{1}{2}\textit{F}_{\mu\lambda}\textit{F}_{\nu}^{ \ \ \lambda}+\frac{1}{4}g_{\mu\nu}m^{2}\Psi^{2}+
\frac{1}{4}g_{\mu\nu}(\partial\Psi)^{2}  &\nonumber \\ -\frac{1}{2}\partial_{\mu}\Psi\partial_{\nu}\Psi  + \frac{1}{4}g_{\mu\nu}\textrm{G}(\Psi)(\partial\alpha-A)^{2} - \frac{1}{2}\textrm{G}(\Psi)(\partial_{\mu}\alpha-A_{\mu})(\partial_{\nu}\alpha-A_{\nu})    =0
\label{einsteineom}
\end{eqnarray}
And finally, the EOM for $\alpha$ is given as
\begin{eqnarray}
\partial_{\mu}\biggl(\sqrt{-g}\textrm{G}(\Psi)(\partial^{\mu}\alpha-A^{\mu})
\biggr)=0
\end{eqnarray}

In the above equations we have explicitly set $q=1$ and $L=1$. We will henceforth consider a particular gauge where $\alpha=0$.
Since we are mostly interested in finding the charged hairy planar black hole solution, we consider the following ansatz for the metric
 which includes the back-reaction of gauge and scalar field
\begin{eqnarray}
\textit{d}s^{2}=-r^2 g(r)e^{2\chi(r)}dt^{2}+r^{2}(dx^{2}+dy^{2})+\frac{dr^{2}}{r^2 g(r)}
\label{matric}
\end{eqnarray}
The Hawking temperature of this black hole is given by:
\begin{equation}
T_{H}=\frac{r^2 g'(r)e^{\chi(r)}}{4\pi}|_{r=r_{h}}
\end{equation}
where $r_h$ is the radius of event horizon, given by the solution of $g(r_h)=0$. For the scalar and the gauge fields, we consider
\begin{equation}
\Psi=\Psi(r),A=\Phi(r)dt
\label{fieldsansatz}
\end{equation}
In the rest of the paper we will work in z=$r_h/r$ coordinate system which is more convenient for numerical calculations. In this coordinate system,
the horizon is located at $z=1$ and the boundary at $z=0$. With our ansatz, the equations of motion reduce to

\begin{eqnarray}
\Psi''+ \Psi '\left(\frac{g'}{g}- \frac{2}{z}+ \chi'\right) + \frac{ \Phi ^2 e^{-2\chi }}{2
   g^2}\frac{d\textrm{G}(\Psi)}{d\Psi} -\frac{m^2 \Psi }{z^2 g} =0
\label{scalareom}
\end{eqnarray}

\begin{eqnarray}
\Phi''- \chi' \Phi' -\frac{\textrm{G}(\Psi)}{z^2 g}\Phi =0
\label{gaugeeom}
\end{eqnarray}

\begin{eqnarray}
g' - \kappa ^2 \left(\frac{z e^{-2 \chi} \textrm{G}(\Psi) \Phi^2}{2 g}+\frac{1}{2} z g
   \Psi'^2+\frac{m^2 \Psi^2}{2 z}+\frac{1}{2} z^3 e^{-2 \chi} \Phi'^2\right)-\frac{3
  g}{z}+\frac{3}{z}=0
\label{tteinsteineom}
\end{eqnarray}

\begin{eqnarray}
 \chi' + \frac{1}{2}z \kappa^2 \Psi'^2 + \frac{\kappa^{2}z e^{-2\chi} \Phi^2\textrm{G}(\Psi)}{2 g^2}=0
\label{rreinsteineom}
\end{eqnarray}

Here, eqs. (\ref{tteinsteineom}) and (\ref{rreinsteineom}) are the $tt$ and $tt-rr$ components of the Einstein's equation. Note that in above equations,
we have explicitly suppressed the z-dependence of our variables, and the prime denotes a derivative with respect to the z-coordinate.
We therefore have four coupled non-linear differential equations which need to be solved with suitable boundary conditions.
At the horizon $(z=1)$, we impose the regularity conditions for $\Phi$ and $\Psi$, where these fields behave as
\begin{equation}
\Phi(1)=0 ,\ \ \Psi'(1)=\frac{m^2\Psi(1)}{g'(1)}
\label{horizon behavior}
\end{equation}
Also, as we have already mentioned $g(1)=0$. The first condition in eq. (\ref{horizon behavior}) is imposed by demanding a finite form for the $U(1)$ current
at the horizon, and the second condition comes from eq. (\ref{scalareom}).

Near the boundary $(z=0)$, these fields asymptote to the following expressions
\begin{eqnarray}
\Phi=\mu-\rho z +... , \ \ \Psi=\Psi_{-}z^{\lambda_{-}} + \Psi_{+}z^{\lambda_{+}} + ... &\nonumber \\
\chi\rightarrow 0, \ \ \ g\rightarrow 1+...
\label{boundar behavior}
\end{eqnarray}
In order to identify the temperature of the boundary theory with the Hawking temperature of the black hole, we impose $\chi\rightarrow0$ at the boundary.
This is the third condition in eq. (\ref{boundar behavior}). Here, $\lambda_{\pm}=\frac{3\pm\sqrt{9+4m^2}}{2}$. In the rest of the paper, we consider a special
case with $m^2=-2$ which also implies $\lambda_{\pm}=2,1$. Having these simple values of $\lambda_{\pm}$ is also an another reason for choosing $m^2=-2$.
Although $m^2$ is negative, it is above the Breitenlohner-Freedman (BF) bound $m^2_{BF}=-9/4$ for four space-time dimensions \cite{Breitenlohner}.

Now using the holographic dictionary which relates a field in the bulk to an operator at the boundary, we identify the leading falloff of the scalar field
$\Psi_{-}$ as the source and the subleading term $\Psi_{+}$  as the scalar operator at the boundary\footnote{Streakily speaking, this operator $\Psi_{+}$ is the expectation value.}.
 \begin{equation}
<O_{2}>\sim{\Psi_{+}}
\label{O2}
\end{equation}
For the mass under consideration $(m^2=-2)$ the roles of $\Psi_{-}$ and $\Psi_{+}$ is interchangeable \textit{i.e} $\Psi_{+}$ can be considered
as the source and $\Psi_{-}$ as the scalar operator, though this scenario is not considered in this paper. In a similar manner $\mu $ and $\rho$ can be
identified as the chemical potential and the charge density of the boundary theory, respectively. In order to break the $U(1)$ symmetry spontaneously,
we set the source term $\Psi_{-}=0$ as the boundary condition.

\begin{figure}[t!]
\begin{minipage}[b]{0.5\linewidth}
\centering
\includegraphics[width=2.7in,height=2.3in]{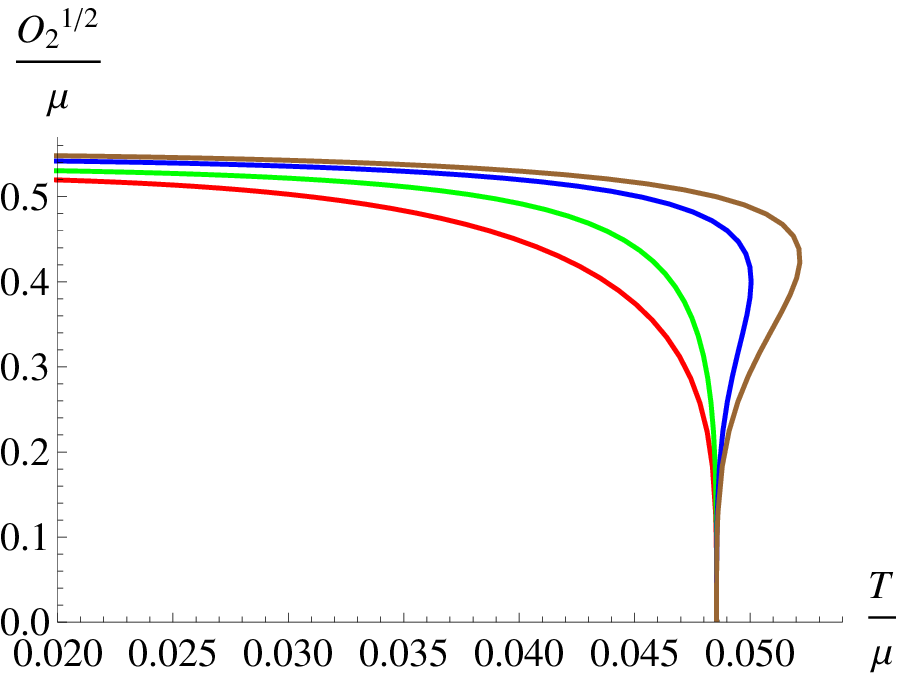}
\caption{Condensate as a function of $T/\mu$ for different values of $\xi$ with fixed $\kappa=0.3$. Here the red, green, blue and brown curves corresponds to $\xi=$ $0$, $0.2$, $0.5$ and $0.7$ respectively.}
\label{O2vsZetaKappaPt3}
\end{minipage}
\hspace{0.4cm}
\begin{minipage}[b]{0.5\linewidth}
\centering
\includegraphics[width=2.7in,height=2.3in]{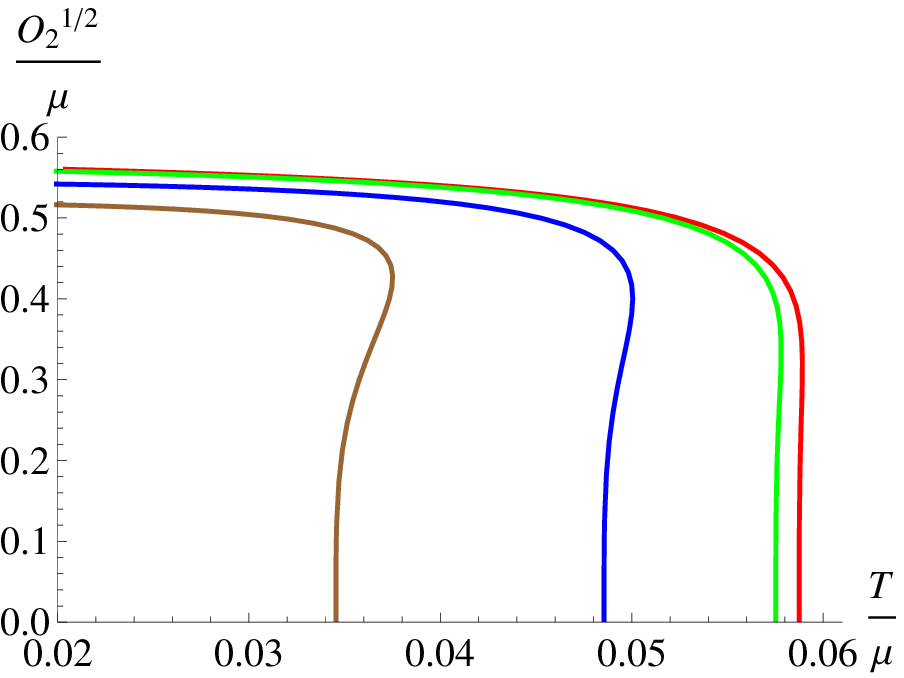}
\caption{Condensate as a function of $T/\mu$ and for different values of $\kappa$ with fixed $\xi=0.5$ .  Here the red, green, blue and brown curves corresponds to $\kappa=$ $10^{-10}$, $0.1$, $0.3$ and $0.5$ respectively.}
\label{O2vsKappaZetaPt5}
\end{minipage}
\end{figure}

For numerical calculations, we will specialize to particular class of forms for $\textrm{G}(\Psi)$, namely
\begin{eqnarray}
\textrm{G}(\Psi)=\Psi^{2}+\xi \Psi^{\theta}
\label{generalized terms}
\end{eqnarray}
To be explicit, we will set $\theta=4$. Other values of $\theta$ in generalized superconductors are also allowed and $\theta=4$ is simply one particular choice.
We have studied other values of $\theta$ as well and have found no qualitative changes in the results. We also mention here that one must choose
$\theta>1$ since otherwise $\Psi$ appears in the denominator in the scalar field equation of motion, and therefore a normal solution $\Psi=0$ will not be allowed.

In figs.(\ref{O2vsZetaKappaPt3}) and (\ref{O2vsKappaZetaPt5}), we have shown the variation of condensate with respect to $T/\mu$.
The red, green, blue and brown curves in fig.(\ref{O2vsZetaKappaPt3}) correspond to $\xi=$ $0$, $0.2$, $0.5$ and $0.7$ respectively with the
backreaction parameter kept fixed at $\kappa=0.3$. For all the cases, we find a critical $T/\mu$ below which the scalar operator $O_2$ becomes
nonzero and develops a non vanishing vacuum expectation value which indicates the spontaneous breaking of global $U(1)$ symmetry at the boundary
and therefore a phase transition to the superconducting phase. Above the critical $T/\mu$, the system is in the normal phase with $O_2=0$.
We find that the phase transition from normal to superconducting phase is present for all $\xi$. However, the order of the phase transition depends
crucially on the magnitude of $\xi$. This can be seen from fig.(\ref{O2vsZetaKappaPt3}) where small $\xi=$$0$, $0.2$ gives second order phase
transitions whereas relatively large $\xi=$$0.5$, $0.7$ gives first order phase transitions. This behavior indicates the existence of a lower cutoff $\xi_c$,
at a particular $\kappa$, above which the transition form normal to superconducting phase is always first order. We have explicitly checked that
for various values of $\kappa$, such a $\xi_c$ always exist. Importantly, The critical $T/\mu$ does not depends on $\xi$. This can be seen
from eq.(\ref{generalized terms}), since at the critical point, the scalar field is small and $\xi$ which comes with higher powers of the scalar field
does not have any effects on the critical $T/\mu$.

Similarly, the variation of condensate with fixed $\xi=0.5$, but for different values of $\kappa$ is shown in fig.(\ref{O2vsKappaZetaPt5}). In this figure, the
red, green, blue and orange curves correspond to $\kappa=$$10^{-10}$, $0.1$ $0.3$ and $0.5$ respectively. We again see the phase transition to the superconducting
phase but now the transition point changes with $\kappa$, namely higher backreaction makes critical $T/\mu$ smaller. This is an important difference
with respect to  fig.(\ref{O2vsZetaKappaPt3}) where critical $T/\mu$ does not depend on $\xi$.

By considering other values of $\theta$, different structure of the phase transition can also be explored.  In particular, by taking $\theta=6$ or $8$,
we find a metastable region with in the superconducting phase. however, the overall results such as the existence of $\xi_c$ etc remain the same for all $\theta$.

\section{Holographic response functions}

Having established the superconducting nature of the boundary theory, we now proceed to calculate its response functions. As mentioned in section 2,
in order to obtain the response functions, we first need to calculate the transverse current-current correlators. In the AdS/CFT correspondence, these
can be computed by looking at the linear response of the superconducting system to the gauge field perturbations, which without any loss of generality
 can be considered as the $x$ component of the gauge field, $A_x$. Since we are working with backreaction, in order to have consistent Einstein and
 Maxwell equations, the gauge field perturbation will have to be supplemented with metric perturbations. We work with metric perturbations
 $g_{xt}(z)\neq0$ and $g_{xy}(z)\neq0$. One can easily check that the independent set $\{A_x(z), g_{xt}(z), g_{xy}(z)\}$  of perturbations are consistent
 with the Einstein as well as the gauge field equation of motion. These are commonly called the vector type
 perturbations.

From now on, we assume a time and momentum dependence of the form $e^{-i\omega t+ iky}$, \textit{i.e}
$$A_x\sim A_x(z) e^{-i\omega t+ iky},  g_{tx}\sim  g_{tx}(z) e^{-i\omega t+ iky},  g_{xy}\sim  g_{xy}(z) e^{-i\omega t+ iky}  $$
It is more convenient to introduce new variables
$$g^x_{t}(z)= z^2 g_{xt}(z), g^x_{y}= z^2 g_{xy}(z)$$
With these redefinition of variables, we use the Maxwell's and Einstein's equations. At the linearised level, the $x$-component of the Maxwell's equations reads
\begin{eqnarray}
A_{x}''+A_{x}'\biggl(\frac{g'}{g}+\chi' \biggr)+A_{x}\biggl(\frac{\omega^2 e^{-2\chi}}{g^2} -\frac{k^2
   }{g}-\frac{\textrm{G}(\Psi)}{z^2 g}\biggr)+\frac{e^{-2 \chi}\Phi' g{^{x}_t}' }{g}=0
\label{Axeom}
\end{eqnarray}
and the $(t,x)$, $(x,y)$ and $(r,x)$-components of Einstein's equation give
\begin{eqnarray}
g{^x_t}'' - g{^x_t}'\biggl(\chi'+ \frac{2}{z}\biggr)+ 2\kappa ^2 z^2 A_x'\Phi'+ \frac{2 \kappa^2 \Phi \textrm{G}(\Psi) A_x}{g(z)} - \frac{k^2 g{^x_t}}{g} - \frac{k \omega g{^x_y}}{g}=0
\label{gxteom}
\end{eqnarray}
\begin{eqnarray}
g{^x_y}''+ g{^x_y}'\biggl(\frac{g'}{g}+\chi' -\frac{2}{z}\biggr) +\frac{\omega^2 e^{-2 \chi} g{^x_y}}{g^2} +\frac{k \omega e^{-2 \chi} g{^x_t}}{g^2}=0
\label{gxyeom}
\end{eqnarray}
\begin{eqnarray}
2 \kappa^2 \omega z^2 A_x \Phi'+ k g e^{2 \chi} g{^x_y}' +\omega
   g{^x_t}'=0
\label{constraineom}
\end{eqnarray}
Again, a prime denotes a derivative with respect to $z$, and we have explicitly suppressed the $z$ dependence of the variables.
Eqs. (\ref{gxteom})-(\ref{constraineom}) are not independent, in particular eq. (\ref{gxteom}) and (\ref{constraineom}) imply eq. (\ref{gxyeom}).
It is clear from the nature of these coupled differential equations that the calculations for response functions including backreaction are complicated
and can be difficult to solve even numerically. However in the probe limit $\kappa\rightarrow 0$, in which the backreaction of gauge and scalar fields
on the space-time geometry can be neglected, metric fluctuations are set to zero and therefore one only needs to consider the $A_x$ equation of motion.
In this simple limit, therefore, the calculations of transverse current-current correlators and hence the response functions become simpler.
This has been done in many previous works, see\cite{Gao}\cite{Subhash}. In this paper, our main aim is to go away from the probe limit, and to numerically calculate
the response functions with backreaction.

Before going into the details of solving these differential equations explicitly, a word about the solution for the normal phase $\Psi=0$.
As we have already mentioned, generally these coupled differential equations are difficult to solve. However when $\Psi=0$, one can construct the
``master variables'', which are linear combinations of the $\{A_x(z), g_{xt}(z), g_{xy}(z)\}$ fields. Written in terms of these master
variables eqs. (\ref{Axeom})-(\ref{constraineom}) reduce to two decoupled set of equations. Although, the construction of master variable is mostly
a trial and error method, it gives us a remarkably easier handle on the analytical as well as the numerical aspects of the perturbations. However,
due to the presence of the $\Psi$ and $\chi$ fields, it is not possible to construct the master variables in the superconducting phase.
Therefore, one needs to resort to a numerical study.

We now proceed to solve eqs. (\ref{Axeom})-(\ref{constraineom}) numerically. We need to apply certain boundary conditions. The natural choice at the
horizon is the infalling wave boundary condition. This is mathematically equivalent to imposing $A_x\propto f(z)^{\frac{-i \omega}{4 \pi T}}$
(and similar expressions for other fields) at the horizon. For the current-current correlator, we also need the on-shell action that gives the finite and the
quadratic function of the boundary values
\begin{eqnarray}
\textit{S}_{on-shell}= \int \biggl(\frac{1}{4z^2\kappa^2}(g{^x_t}g{^x_t}'-g g{^x_y}g{^x_y}')-\frac{g}{2}(A_{x}A_{x}'-\rho g{^x_t})\biggr)^{z=1}_{z=0}\,
\label{sonshell}
\end{eqnarray}
where we have removed all the contact terms. \footnote{The full $\textit{S}_{on-shell}$ also contains terms like $g{^x_t}g{^x_t}/z^3$, $g{^x_y}g{^x_y}/z^3$
\textit{etc} which gives divergent contribution to  the correlators. These divergences can be subtracted by a holographic renormalization procedure,
which in our set up corresponds to adding a constant term to the correlators. We will fix the constant term by requiring
$Re(\varepsilon(\omega))\rightarrow 1$ at large frequencies.} One notices that the first term in above equation is only due to the metric fluctuations and
would be absent in the probe limit. This term is important for the negative refraction to occur as it introduces a diffusive pole
in the current-current correlator.\footnote{See the discussion in section $3$ of \cite{Amariti}.}

Now we proceed to the calculation for the transverse current-current correlator. Generally, in AdS/CFT correspondence, we obtain this correlator by
taking the second derivative of $\textit{S}_{on-shell}$ with respect to the boundary value of $A_x$ fluctuations \cite{Son}. However, due to the
coupling of $A_x$ with metric fluctuations, this procedure is difficult to implement numerically. Consequently, we will use another elegant technique
developed in \cite{Amariti}. Below, we will briefly review the salient features of this technique.

Let us first record the near boundary $z=\alpha$ expansion
\begin{eqnarray}
& A_{x}'(\alpha)=\frac{1}{A_{xx}}\biggl((G_{xtx}-\rho A_{xx})g{^x_t}(0)-G_{xyx}g{^x_y}(0)+G_{xx}A_x(0) \biggr) &\nonumber \\
& g{^x_t}'(\alpha)=\frac{1}{A_{xtxt}}\biggl(G_{xtxt}g{^x_t}(0)-G_{xtxy}g{^x_y}(0)+G_{xtx}A_x(0) \biggr) &\nonumber \\
& g{^x_y}'(\alpha)=\frac{1}{A_{xyxy}}\biggl(-G_{xtxy}g{^x_t}(0)+G_{xyxy}g{^x_y}(0)-G_{xyx}A_x(0) \biggr)
\label{boundayAxexp}
\end{eqnarray}
Where $A_{xtxt}=-A_{xyxy}=1/(4 \kappa^2 \alpha^2)$. $A_x(0)$, $g{^x_t}(0)$ and $g{^x_y}(0)$ are the boundary values of
$A_x(z)$, $g{^x_t}(z)$ and $g{^x_y}(z)$ respectively, and $G_{xx}$, $G_{xtxt}$, $G_{xyxy}$ \textit{etc} are the correlators for the current and
the energy-momentum tensor. The expansions in eq. (\ref{boundayAxexp}) directly follow from $\textit{S}_{on-shell}$ in eq. (\ref{sonshell}).
For the normal phase, one can explicitly check validity of the expansion used in eq. (\ref{boundayAxexp})\cite{Ge}.

The main idea in \cite{Amariti} is to first solve eqs. (\ref{Axeom}), (\ref{gxteom}), (\ref{gxyeom})  simultaneously,
and then treat eq. (\ref{constraineom}) separately  as the constraint equation on the various correlators. The constraints on the
correlators from eq. (\ref{constraineom}) can be seen as follows. If we multiply eq. (\ref{constraineom}) by $A_x(0)$
(similarly with $g{^x_t}(0)$, $g{^x_y}(0)$) and take double derivative with respect to $A_x(0)$ ($g{^x_t}(0)$, $g{^x_y}(0)$), we get
\begin{eqnarray}
& \ \ \ \ \ \ \ \omega G_{xtx}+k G_{xyx}+2\kappa^2\omega\Phi'(z)z^2A_{xtxt}=0 &\nonumber \\
&  \omega G_{xtxt}+k G_{xtxy}=0 &\nonumber \\
& \omega G_{xtxy}+k G_{xyxy}=0
\label{constrainGxx}
\end{eqnarray}
We have explicitly used eq. (\ref{boundayAxexp}) to derive the above equations. One can see that eq. (\ref{constrainGxx}) is the constraint
equation on the various correlators and that all the correlators are not independent. These three constraint equations together with
eq. (\ref{Axeom}), (\ref{gxteom}), (\ref{gxyeom}) imply that we have a total of six equations which need to be solved in order to calculate the six
correlators. \footnote{We are assuming the $G_{ij}=G_{ji}$ symmetry.}
Now, in order to obtain the response functions, we also need to expand the correlators and perturbation fields in powers of $k$.
Using eq.(\ref{constraineom}) and Lorentz invariance, one finds that the only consistent expansion in a series of $k$ reads (upto $O(k^2)$ terms) :
\begin{eqnarray}
A_x=A_{x0}+ k^2 A_{x2}, \ g{^x_t}=g{^x_{t0}}+ k^2 g{^x_{t2}}, \ g{^x_y}=k g{^x_{y1}}+ k^2 g{^x_{y2}} & & \nonumber \\
G_{xx}=G_T^{(0)}+k^2G_T^{(2)}, \ G_{xtxt}=G_{xtxt0}+k^2G_{xtxt2}, \ G_{xyxy}=G_{xyxy0}+k^2G_{xyxy2}
\label{expGxx}
\end{eqnarray}
and similar expansions for $G_{xtxy}$, $G_{xtx}$ and $G_{xyx}$ can be obtained. After implementing the above procedure, the computation of the response functions
now boils down to numerically finding $G_T^{(0)}$ and $G_T^{(2)}$.

Before proceeding, we remind the reader of the caveat  mentioned in the introduction, namely the absence of a dynamical photon at the boundary.
In the AdS/CFT correspondence, the gauge symmetry in the bulk corresponds to a global symmetry at the boundary. Therefore, strictly speaking there is
no dynamical photon at the boundary. However, one can weakly gauge the symmetry to a dynamical photon at the boundary \cite{Hartnoll1}.
In other words, we are calculating the response function of the boundary systems from AdS/CFT correspondence and introducing the coupling between the system and the photon by hand in a perturbative manner \footnote{see discussion below eq.(15) of \cite{Policastro} in order to see that we are indeed
considering the leading effects of the dynamical photon weakly coupled to the boundary system.}. This is standard in the literature\cite{Policastro}\cite{Hartnoll1}
and we are also adopting the same procedure.

Before we present our numerical results, we state the differences in our setup with the model considered in \cite{Amariti}.
We emphasize here that we investigating the response functions of generalized holographic superconductors in $(2+1)$ dimensions
with non minimal interaction between the scalar and the gauge field. This is significantly different from the holographic superconductor considered in \cite{Amariti} with
minimal interaction in $(3+1)$ dimension and the model considered here has
far richer phase structure.

\begin{figure}[t!]
\begin{minipage}[b]{0.5\linewidth}
\centering
\includegraphics[width=2.5in,height=2.0in]{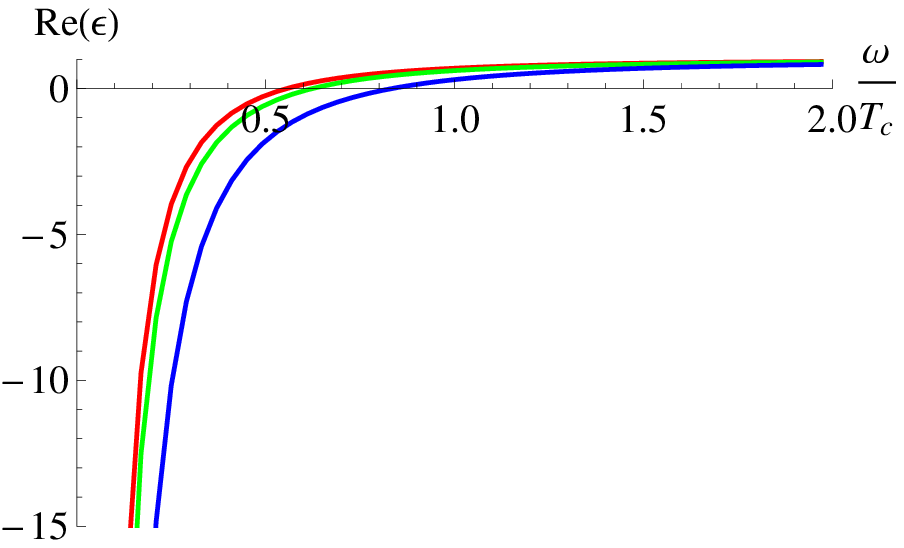}
\caption{$Re(\varepsilon)$ as a function of $\omega/T_c$ for different values of $\kappa$ with fixed $\xi=0.2$ at $T=0.5 T_c$. Here the red, green and blue curves correspond to $\kappa=$ $0.1$, $0.3$ and $0.5$ respectively.}
\label{ReEpsilonVsKappaZetaPt2TPt5Tc}
\end{minipage}
\hspace{0.4cm}
\begin{minipage}[b]{0.5\linewidth}
\centering
\includegraphics[width=2.5in,height=2.0in]{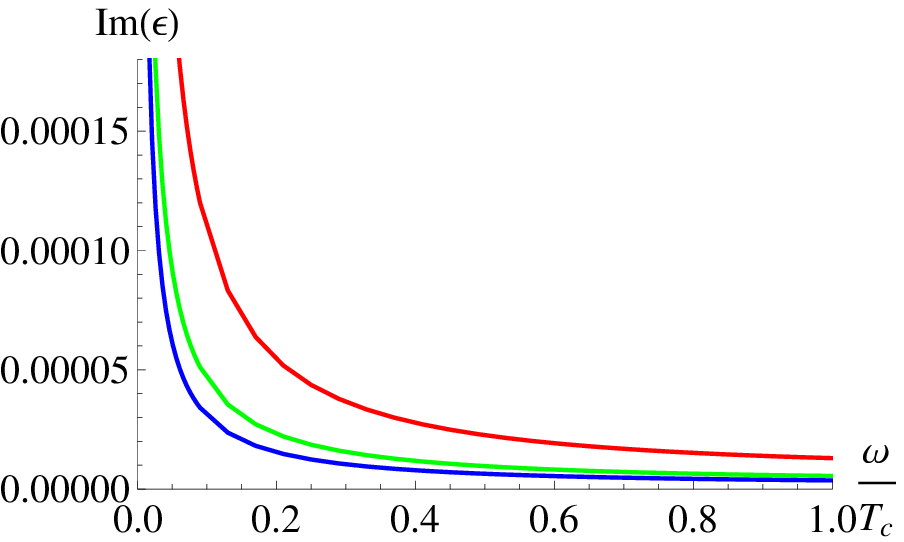}
\caption{$Im(\varepsilon)$ as a function of $\omega/T_c$ for different values of $\kappa$ with fixed $\xi=0.2$ at $T=0.5 T_c$. Here the red, green and blue curves correspond to $\kappa=$ $0.1$, $0.3$ and $0.5$ respectively.}
\label{ImEpsilonVsKappaZetaPt2TPt5Tc}
\end{minipage}
\end{figure}

\begin{figure}[t!]
\begin{minipage}[b]{0.5\linewidth}
\centering
\includegraphics[width=2.5in,height=2.0in]{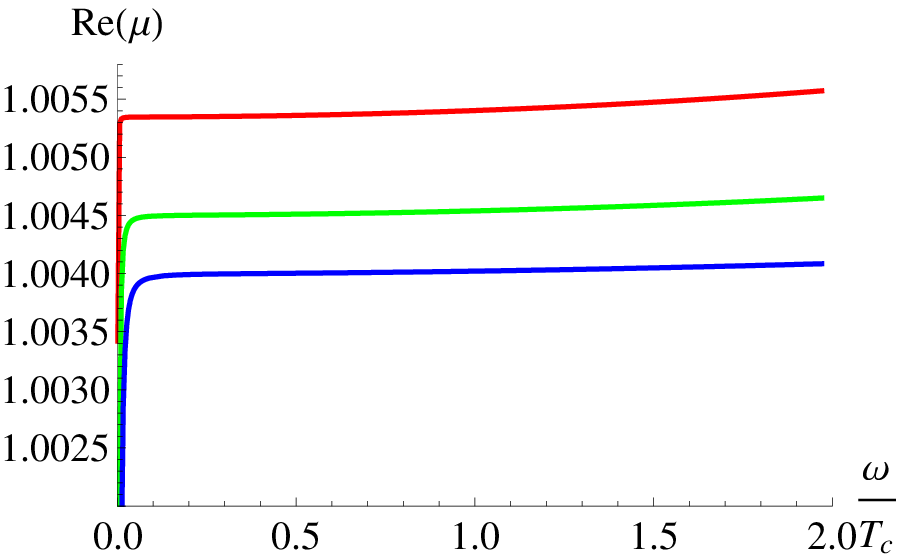}
\caption{$Re(\mu)$ as a function of $\omega/T_c$ for different values of $\kappa$ with fixed $\xi=0.2$ at $T=0.5 T_c$. Here the red, green and blue curves correspond to $\kappa=$ $0.1$, $0.3$ and $0.5$ respectively.}
\label{ReMuVsKappaZetaPt2TPt5Tc}
\end{minipage}
\hspace{0.4cm}
\begin{minipage}[b]{0.5\linewidth}
\centering
\includegraphics[width=2.5in,height=2.0in]{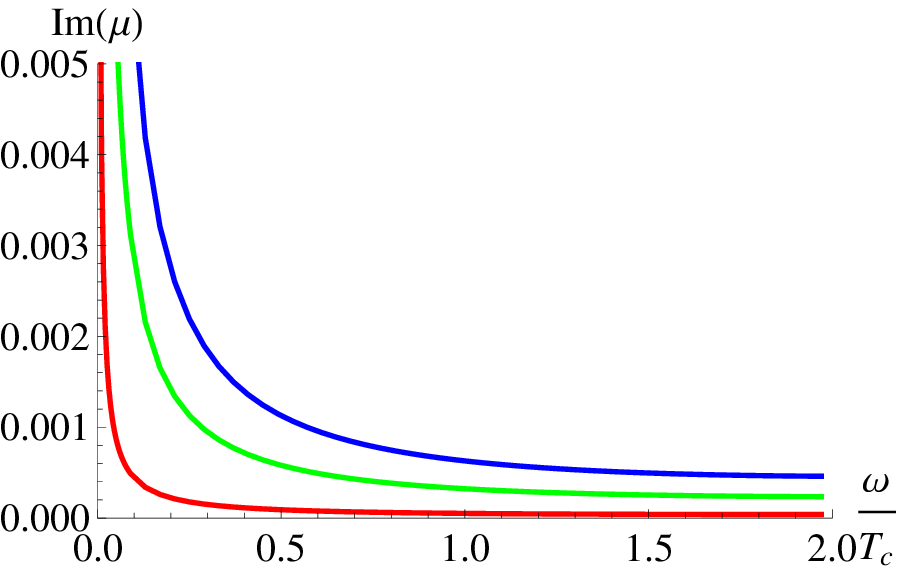}
\caption{$Im(\mu)$ as a function of $\omega/T_c$ for different values of $\kappa$ with fixed $\xi=0.2$ at $T=0.5 T_c$. Here the red, green and blue curves correspond to $\kappa=$ $0.1$, $0.3$ and $0.5$ respectively.}
\label{ImMuVsKappaZetaPt2TPt5Tc}
\end{minipage}
\end{figure}

Now we will present our numerical results for the optical properties of the boundary superconducting system. Here, we will present the results with nonzero $\kappa$, and
the results in the probe limit ($\kappa=0$) can be found in \cite{Subhash}. In figs.(\ref{ReEpsilonVsKappaZetaPt2TPt5Tc}), (\ref{ImEpsilonVsKappaZetaPt2TPt5Tc}),
(\ref{ReMuVsKappaZetaPt2TPt5Tc}) and (\ref{ImMuVsKappaZetaPt2TPt5Tc}), we have shown the variation of $Re(\varepsilon)$, $Im(\varepsilon)$,
$Re(\mu)$ and $Im(\mu)$ respectively with respect to $\omega/T_c$. In all these figures, the red, green and blue curves correspond to
$\kappa=0.1$, $0.3$ and $0.5$ respectively and we have fixed the temperature at $T=0.5T_c$ and $\xi=0.2$.
Our results in 4D AdS-Schwarzschild background with backreaction, for the electric permittivity and the effective magnetic permeability are
qualitatively similar to those obtained in \cite{Amariti} for holographic superconductors in $5D$ AdS-Schwarzschild background.
However, here we considered generalized holographic superconductors with nonzero $\xi$ which is distinct from the case studied in \cite{Amariti}.
The $Re(\varepsilon)$ at very low frequency becomes large negative and approaches to unity towards higher frequency. The behavior of $Re(\varepsilon)$
at large frequency is expected on physical grounds : the system does not have the time to respond to the rapid variation of the external
perturbation and therefore behaves like vacuum. The $Im(\varepsilon)$ has a pole at $\omega=0$ and is always positive. Using the standard
result which relates $\varepsilon(\omega)$ to the optical conductivity $\sigma(\omega)$ as $\varepsilon(\mu)=1+4 i\pi \sigma(\omega)/\omega $,
we find that $Re(\sigma)$ has a delta function singularity at zero frequency, which is one of the main characteristic features of superconductivity.
The results for electrical permittivity with backreaction are similar to the ones found in the probe limit in \cite{Subhash}.

\begin{figure}[t!]
\begin{minipage}[b]{0.5\linewidth}
\centering
\includegraphics[width=2.5in,height=2.0in]{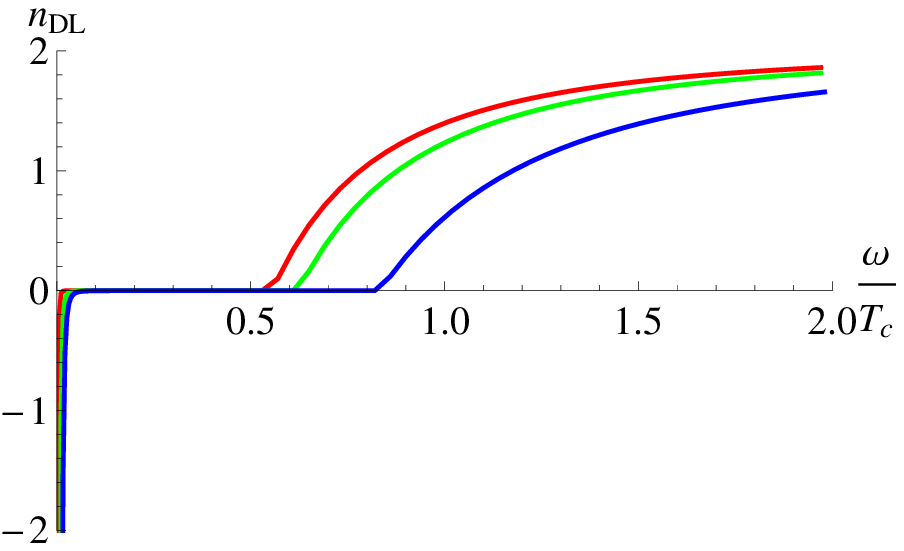}
\caption{$n_{DL}$ as a function of $\omega/T_c$ for different values of $\kappa$ with fixed $\xi=0.2$ at $T=0.5 T_c$. Here the red, green and blue curves correspond to $\kappa=$ $0.1$, $0.3$ and $0.5$ respectively.}
\label{NDLVsKappaZetaPt2TPt5Tc}
\end{minipage}
\hspace{0.4cm}
\begin{minipage}[b]{0.5\linewidth}
\centering
\includegraphics[width=2.5in,height=2.0in]{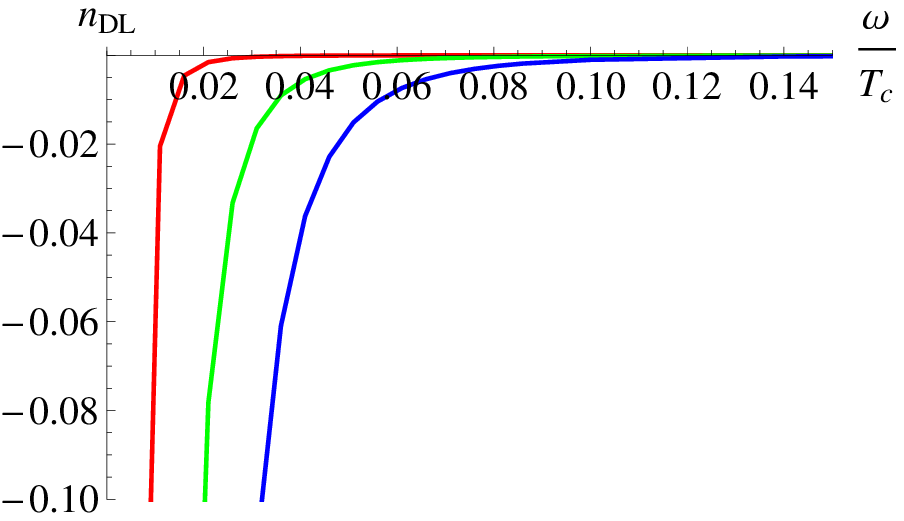}
\caption{Low frequency behavior of $n_{DL}$ as a function of $\omega/T_c$ for different values of $\kappa$ with fixed $\xi=0.2$ at $T=0.5 T_c$. Here the red, green and blue curves correspond to $\kappa=$ $0.1$, $0.3$ and $0.5$ respectively.}
\label{NDLsmallVsKappaZetaPt2TPt5Tc}
\end{minipage}
\end{figure}

Two important contrasts with the results in the probe limit case are: the positivity of $Im(\mu)$ and the presence of pole at $\omega=0$ in $Im(\mu)$.
This is shown in fig.(\ref{ImMuVsKappaZetaPt2TPt5Tc}). Generally, in the probe limit $Im(\mu)$ is always negative and is zero at $\omega=0$.
The appearance of a new pole in $Im(\mu)$ at $\omega=0$ can be attributed to the metric fluctuations in the backreacted case.
The metric fluctuations introduce a new pole (diffusive pole) in the imaginary part of $G_T^{(2)}$, which in turn introduces a pole in the $Im(\mu)$\cite{Matsuo}. This can be seen from eq.(\ref{sonshell}) where the first term leads to a diffusive pole with backreaction
and is absent in the probe limit. We will shortly see that this pole in $Im(\mu)$ also greatly enhances the possibility of negative refraction below a
certain critical frequency. The negativity of $Im(\mu)$ which generally occur in the probe limit, as already mentioned in section $2$, normally implies the
breakdown of $\varepsilon-\mu$ approach, though the $\mu$ appearing in eq.(\ref{epsilonmu}) is an effective magnetic permeability and is not an observable.
However, we see in fig.(\ref{ImMuVsKappaZetaPt2TPt5Tc}) that the backreaction makes the $Im(\mu)$ positive.

Our main purpose here is to calculate $n_{DL}$ and its dependence on $\kappa$. This is shown in figure in fig.(\ref{NDLVsKappaZetaPt2TPt5Tc}),
where we have used the same color coding as in fig.(\ref{ReEpsilonVsKappaZetaPt2TPt5Tc}). We see that the superconducting system makes the
transition from positive $n_{DL}$ to negative $n_{DL}$ as we decrease $\omega$. It can be clearly seen in fig.(\ref{NDLsmallVsKappaZetaPt2TPt5Tc}),
where we focus on the low frequency behavior of $n_{DL}$. This implies the presence of a cutoff $\omega_c$ below which $n_{DL}$ is negative.
Above $\omega_c$, $n_{DL}>0$ and we have positive refraction. The magnitude of cutoff $\omega_c$ increases with increase in backreaction,
which implies that the superconducting phase can support negative refraction for relatively higher frequencies with higher backreaction. An important difference
with respect to the probe limit case now is that the magnitude of $n_{DL}$ is large below $\omega_c$.

\begin{figure}[t!]
\begin{minipage}[b]{0.5\linewidth}
\centering
\includegraphics[width=2.5in,height=2.0in]{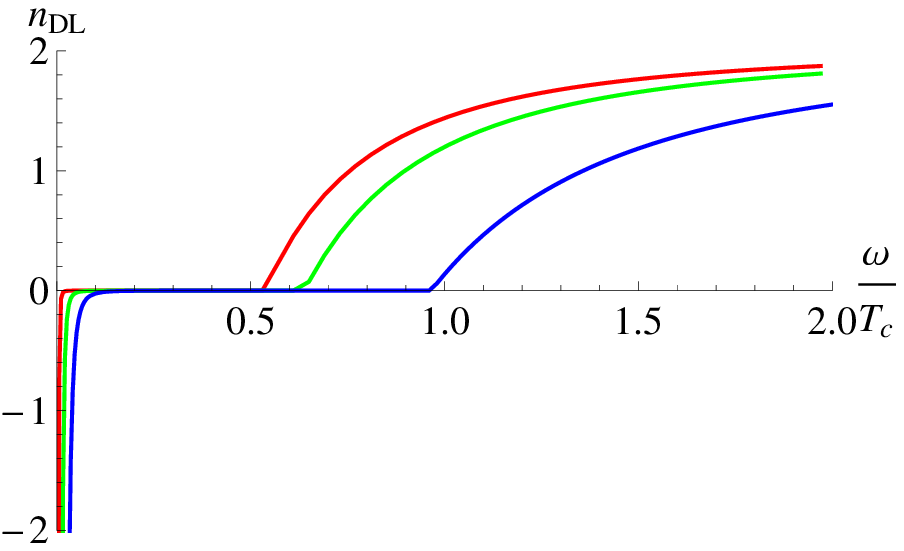}
\caption{$n_{DL}$ as a function of $\omega/T_c$ for different values of $\kappa$ with fixed $\xi=0$ at $T=0.5 T_c$. Here the red, green and blue curves correspond to $\kappa=$ $0.1$, $0.3$ and $0.5$ respectively.}
\label{NDLVsKappaZeta0TPt5Tc}
\end{minipage}
\hspace{0.4cm}
\begin{minipage}[b]{0.5\linewidth}
\centering
\includegraphics[width=2.5in,height=2.0in]{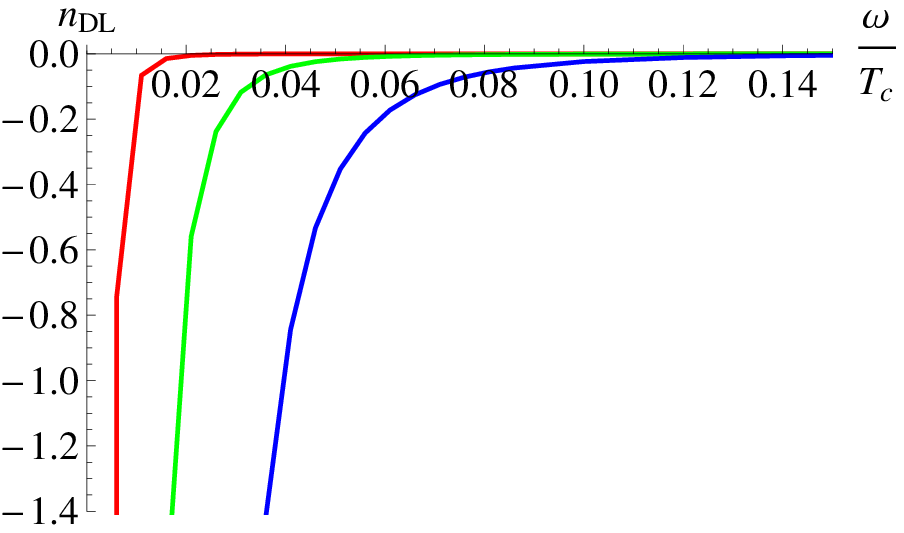}
\caption{Low frequency behavior of $n_{DL}$ as a function of $\omega/T_c$ for different values of $\kappa$ with fixed $\xi=0$ at $T=0.5 T_c$. Here the red, green and blue curves correspond to $\kappa=$ $0.1$, $0.3$ and $0.5$ respectively.}
\label{NDLsmallVsKappaZeta0TPt5Tc}
\end{minipage}
\end{figure}

\begin{figure}[t!]
\begin{minipage}[b]{0.5\linewidth}
\centering
\includegraphics[width=2.5in,height=2.0in]{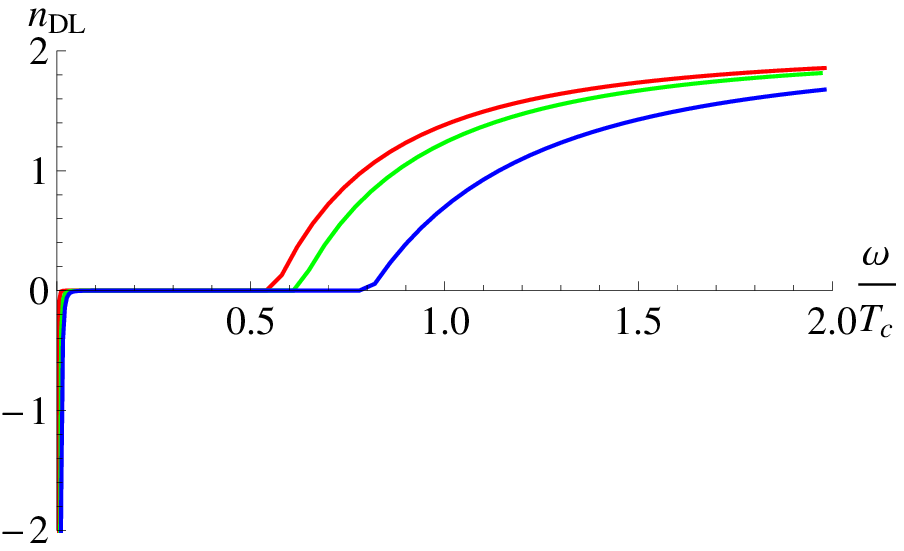}
\caption{$n_{DL}$ as a function of $\omega/T_c$ for different values of $\kappa$ with fixed $\xi=0.5$ at $T=0.5 T_c$. Here the red, green and blue curves correspond to $\kappa=$ $0.1$, $0.3$ and $0.5$ respectively.}
\label{NDLVsKappaZetaPt5TPt5Tc}
\end{minipage}
\hspace{0.4cm}
\begin{minipage}[b]{0.5\linewidth}
\centering
\includegraphics[width=2.5in,height=2.0in]{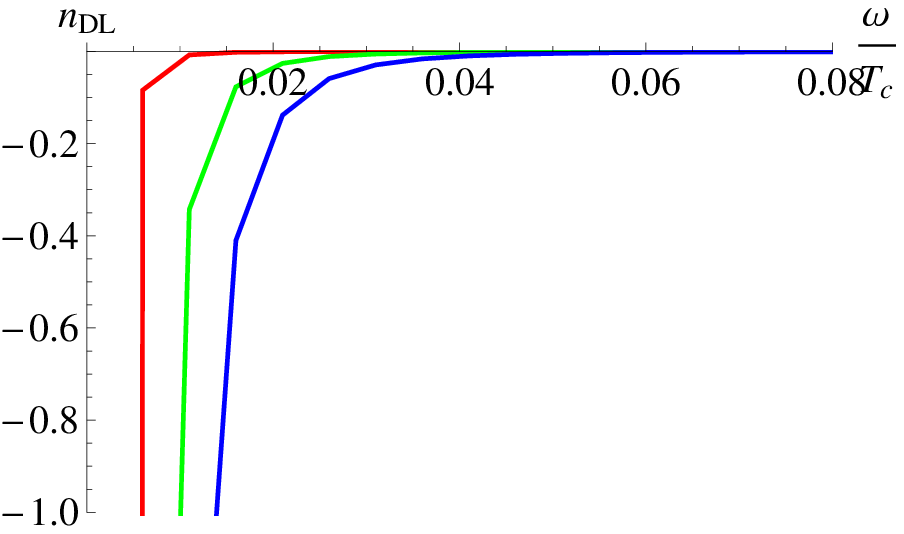}
\caption{Low frequency behavior of $n_{DL}$ as a function of $\omega/T_c$ for different values of $\kappa$ with fixed $\xi=0.5$ at $T=0.5 T_c$. Here the red, green and blue curves correspond to $\kappa=$ $0.1$, $0.3$ and $0.5$ respectively.}
\label{NDLsmallVsKappaZetaPt5TPt5Tc}
\end{minipage}
\end{figure}

We have also computed $n_{DL}$ for other values of the generalized parameter $\xi$ as well. The results are shown in
figs.(\ref{NDLVsKappaZeta0TPt5Tc}) and (\ref{NDLVsKappaZetaPt5TPt5Tc}), where we have plotted $n_{DL}$ for $\xi=0$ and $\xi=0.5$ respectively
at temperature $T=0.5T_c$. The low frequency behavior of these figures are shown in figs.(\ref{NDLsmallVsKappaZeta0TPt5Tc}) and
(\ref{NDLsmallVsKappaZetaPt5TPt5Tc}) respectively. As expected, the essential features of our analysis are similar to the $\xi= 0.2$ case.
We also note that, unlike the case where we varied $\kappa$, the transition from positive refraction to negative refraction with frequency is almost
independent of $\xi$. We have checked this for several values of $\xi$.

\begin{figure}[t!]
\begin{minipage}[b]{0.5\linewidth}
\centering
\includegraphics[width=2.5in,height=2.0in]{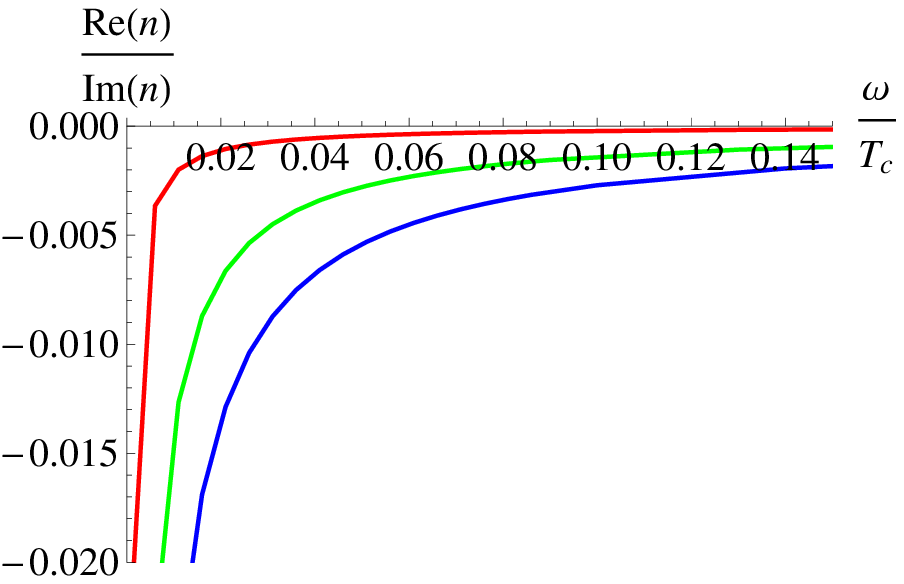}
\caption{Propagation to dissipation ratio as a function of $\omega/T_c$ for different values of $\kappa$ with fixed $\xi=0.2$ at $T=0.5 T_c$.
The red, green and blue curves correspond to $\kappa=$ $0.1$, $0.3$ and $0.5$ respectively.}
\label{RenByImnsKappaZetaPt2TPt5Tc}
\end{minipage}
\hspace{0.4cm}
\begin{minipage}[b]{0.5\linewidth}
\centering
\includegraphics[width=2.5in,height=2.0in]{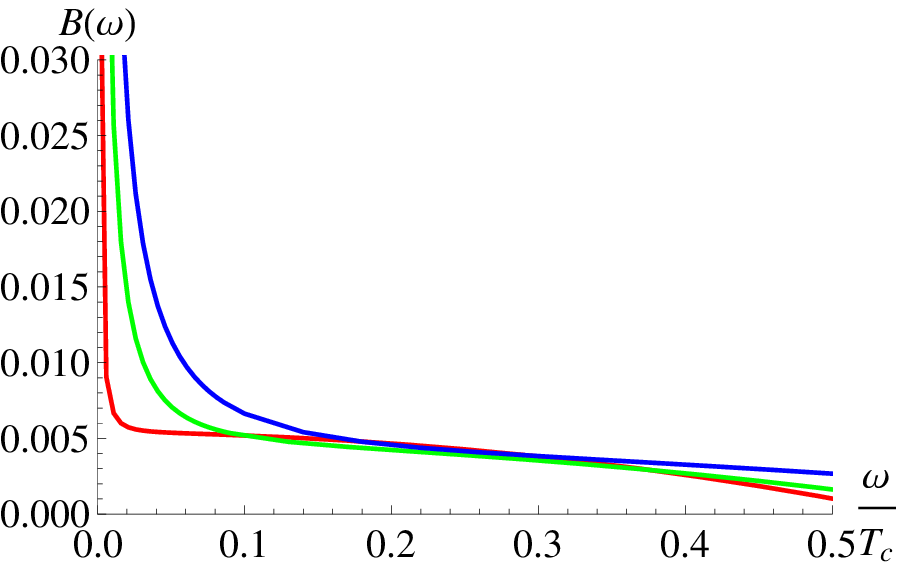}
\caption{$B(\omega)$ as a function of $\omega/T_c$ for different values of $\kappa$ with fixed $\xi=0.2$ at $T=0.5 T_c$. The
red, green and blue curves correspond to $\kappa=$ $0.1$, $0.3$ and $0.5$ respectively.}
\label{ValidityVsKappaZetaPt2TPt5Tc}
\end{minipage}
\end{figure}

It is also important to analyse the propagation $Re(n)$ and the dissipation $Im(n)$ of electromagnetic waves in systems which supports negative refraction.
An important feature of negative refraction is the relative opposite sign between $Re(n)$ and $Im(n)$ \cite{McCall}. Indeed, we find opposite sign between
$Re(n)$ and $Im(n)$ and this is shown in fig.(\ref{RenByImnsKappaZetaPt2TPt5Tc}). We see that $Re(n)/Im(n)<0$ in the frequency range where $n_{DL}$
is negative. Note that, the magnitude of $Re(n)/Im(n)$ is small within the negative refraction frequency range and indicates large dissipation in the system.
However, the higher values of backreaction does enhance the magnitude of $Re(n)/Im(n)$, thereby increasing the propagation. Unfortunately the propagation, on the other hand, decreases with higher values of $\xi$. However, small magnitude of $Re(n)/Im(n)$ a generic result in all holographic setups, and is not unexpected.

For our whole analysis to be trustworthy, it is essential to verify the validity of the expansion used in eq.(\ref{GTexp}).
This is shown in fig.(\ref{ValidityVsKappaZetaPt2TPt5Tc}) where the notation $|\frac{k^2G^{(2)}_T(\omega)}{G^{(0)}_T(\omega)}|=B(\omega)$ has been used.
We see that within the negative refraction frequency range $B(\omega)<1$ and therefore the expansion is reliable. We have checked for other
values of $\xi$ and $\kappa$ as well that $B(\omega)$ is always less than $1$ in the small frequency region.

\begin{figure}[t!]
\begin{minipage}[b]{0.5\linewidth}
\centering
\includegraphics[width=2.5in,height=2.0in]{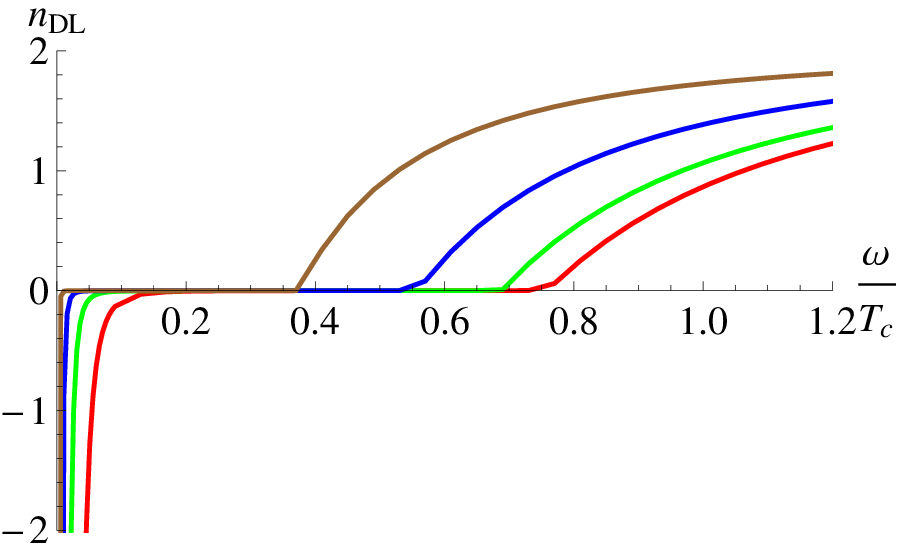}
\caption{$N_{DL}$ as a function of $\omega/T_c$ for different temperatures with fixed $\xi=0$ and $\kappa=0.3$. Here the red, green, blue and brown curves correspond to $T/T_c=$ $0.8$, $0.6$, $0.4$ and $0.2$ respectively.}
\label{NDLVsTempKappaPt3Zeta0}
\end{minipage}
\hspace{0.4cm}
\begin{minipage}[b]{0.5\linewidth}
\centering
\includegraphics[width=2.5in,height=2.0in]{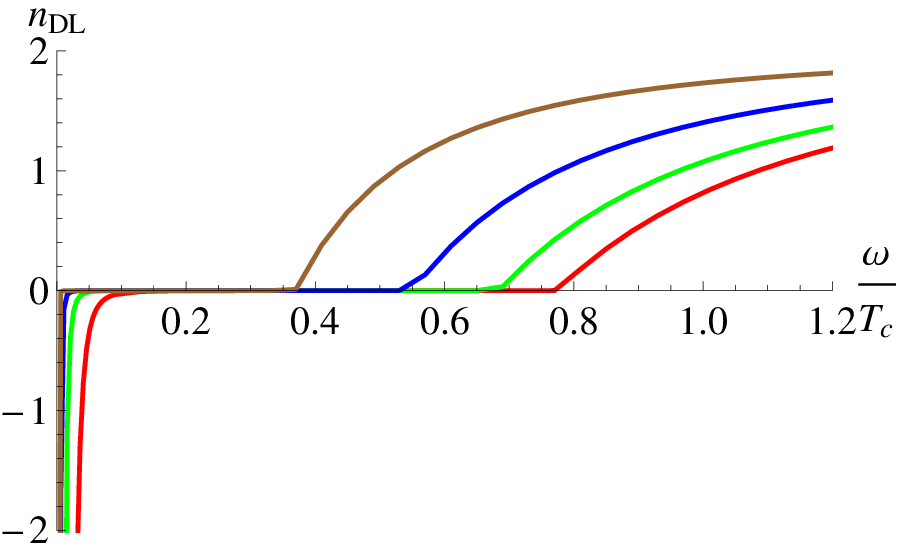}
\caption{$N_{DL}$ as a function of $\omega/T_c$ for different temperatures with fixed $\xi=0.2$ and $\kappa=0.3$. Here the red, green, blue
and brown curves correspond to $T/T_c=$ $0.8$, $0.6$, $0.4$ and $0.2$ respectively.}
\label{NDLVsTempKappaPt3ZetaPt2}
\end{minipage}
\end{figure}

In order to complete our analysis, in figs.(\ref{NDLVsTempKappaPt3Zeta0}) and (\ref{NDLVsTempKappaPt3ZetaPt2}) we have shown
the variation of $n_{DL}$ with $\omega$ for different temperatures. In both these figures red, green, blue and brown curves correspond
to $T/T_c=$ $0.8$, $0.6$, $0.4$ and $0.2$ respectively. We find that negative refraction is present for all temperatures. This is another
distinct result from the probe limit case where $n_{DL}$ was found to be negative only within a window of temperatures.

\begin{figure}[t!]
\begin{minipage}[b]{0.5\linewidth}
\centering
\includegraphics[width=2.5in,height=2.0in]{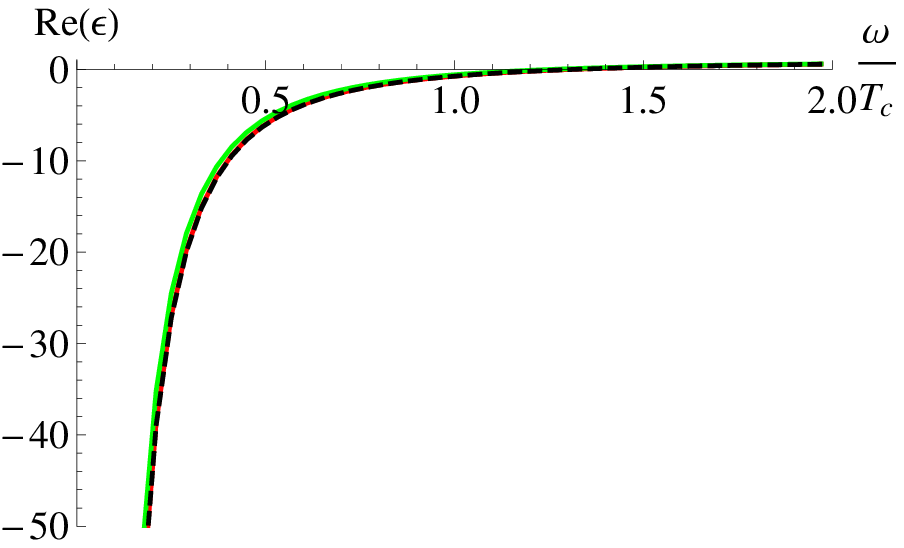}
\caption{$Re(\varepsilon)$ as a function of $\omega/T_c$ with fixed $\xi=0.7$ and $\kappa=0.5$. Here the red,
green and dashed black curves correspond to metastable phase, superconducting phase and normal phase  at $T=1.06T_c$ respectively.}
\label{ReEpsilonKappaPt5ZetaPt7T1Pt06TcPhaseTransition}
\end{minipage}
\hspace{0.4cm}
\begin{minipage}[b]{0.5\linewidth}
\centering
\includegraphics[width=2.5in,height=2.0in]{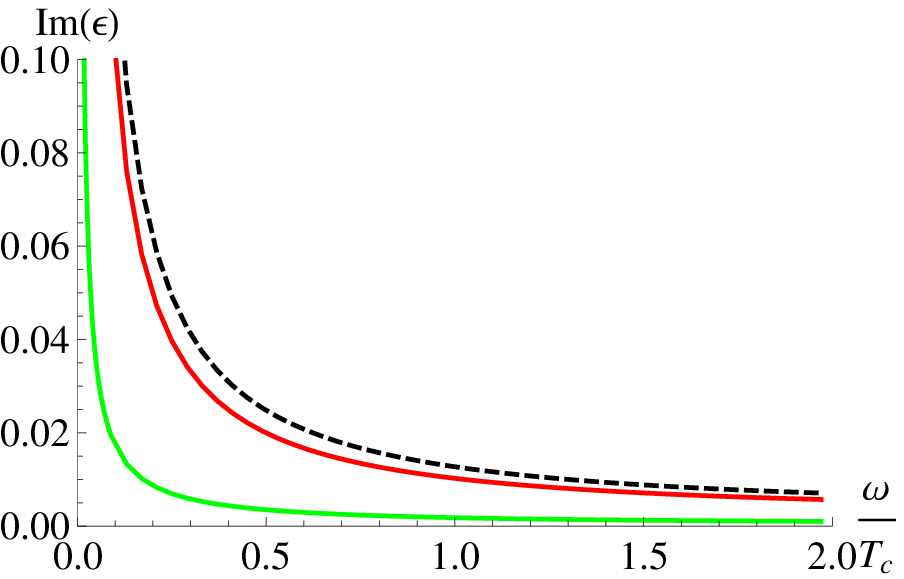}
\caption{$Im(\varepsilon)$ as a function of $\omega/T_c$ with fixed $\xi=0.7$ and $\kappa=0.5$. Here the red,
green and dashed black curves correspond to metastable phase, superconducting phase and normal phase  at $T=1.06T_c$ respectively.}
\label{ImEpsilonKappaPt5ZetaPt7T1Pt06TcPhaseTransition}
\end{minipage}
\end{figure}

\begin{figure}[t!]
\begin{minipage}[b]{0.5\linewidth}
\centering
\includegraphics[width=2.5in,height=2.0in]{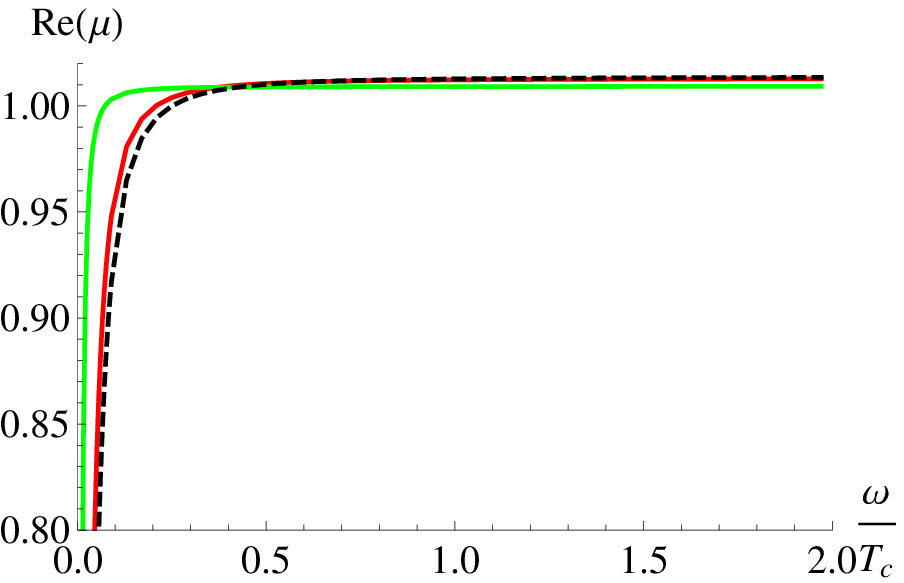}
\caption{$Re(\mu)$ as a function of $\omega/T_c$ with fixed $\xi=0.7$ and $\kappa=0.5$. Here the red, green and dashed black curves correspond to metastable phase, superconducting phase and normal phase  at $T=1.06T_c$ respectively.}
\label{ReMuKappaPt5ZetaPt7T1Pt06TcPhaseTransition}
\end{minipage}
\hspace{0.4cm}
\begin{minipage}[b]{0.5\linewidth}
\centering
\includegraphics[width=2.5in,height=2.0in]{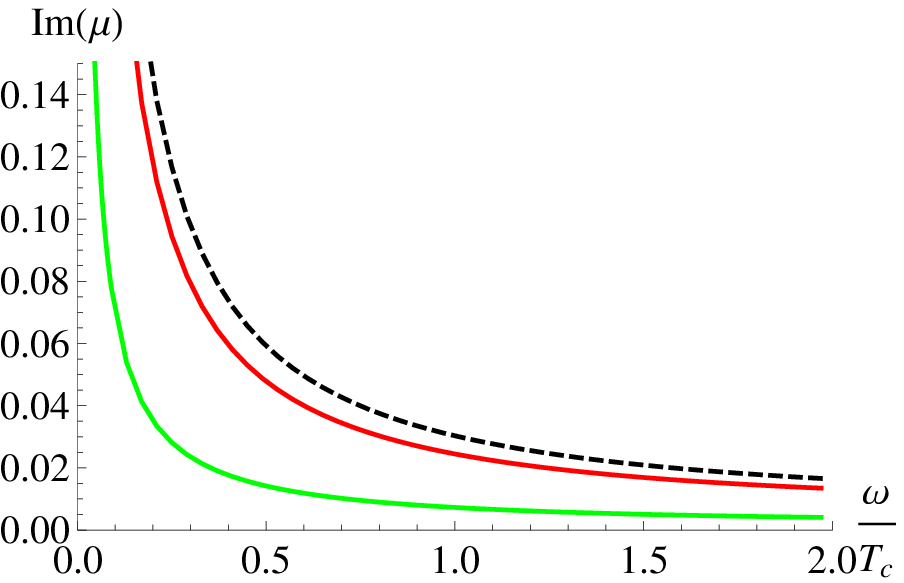}
\caption{$Im(\mu)$ as a function of $\omega/T_c$ with fixed $\xi=0.7$ and $\kappa=0.5$. Here the red, green and dashed black curves correspond to metastable phase, superconducting phase and normal phase  at $T=1.06T_c$ respectively.}
\label{ImMuKappaPt5ZetaPt7T1Pt06TcPhaseTransition}
\end{minipage}
\end{figure}

Finally, it is phenomenologically important to contrast the behavior of the response functions in the normal and the superconducting phases
{\it above} criticality, as this might be experimentally relevant. We do this in figs.(\ref{ReEpsilonKappaPt5ZetaPt7T1Pt06TcPhaseTransition}) -
(\ref{ImMuKappaPt5ZetaPt7T1Pt06TcPhaseTransition}) where we show the behaviour of the permittivity and permeability as a function of
the frequency for $T/T_c = 1.06$, for the stable, metastable and superconducting phases (fig.(\ref{O2vsZetaKappaPt3})). These should be taken as predictions from
holography that can possibly have experimental relevance. For example, for $2+1$ dimensional systems that show a first order transition from the normal
to the superconducting phase, our results suggest that the imaginary part of the permittivity is always smaller in the superconducting phase compared
with the normal phase. Although we are not aware of literature dealing with such systems, testable predictions might be obtained from our
analysis for futuristic experiments.

Before we end this section, we point out that there are a few curious similarities in the response functions of our holographic setup with those obtained from the
Drude model. If we naively compare our results for electric permittivity with the results of Drude model, we find that the parameter $\xi$ naively plays the role
of the inverse of relaxation time $\gamma$ in the Drude picture, and that $\kappa$ is  playing a role analogous to the plasma frequency $\omega_p$. This can be
seen by plotting the $Re(\varepsilon)$ and the $Im(\varepsilon)$ with respect to $\gamma$ and $\omega_p$ in the Drude model, which shows same
qualitative behavior for $\varepsilon$ as we obtained in our holographic setup. This naive identification of $\xi$ with $\gamma$ is further reinforced by
noticing that higher $\xi$, like higher $\gamma$, increases the dissipation in the system. Also in our model, the frequency at which $Re(\varepsilon)=0$ is
independent of $\xi$ but depends on $\kappa$ just like in the Drude model with analogous parameters. We should emphasize here that the Drude model is
not the correct model to describe superconductivity. It captures only a part of the superconductivity characteristics, namely infinite DC conductivity, and cannot
for example explain Meissner effect - which in fact is the essence of superconductivity. However, it is the conductivity which is directly related to the
permittivity and therefore we, in some approximation, can consider the Drude mode to describe these response functions in the superconductors.
 Although the naive identifications of $\xi$ with $\gamma$ and $\kappa$ with $\omega_p$ look reasonable, we make no claims beyond the statement that
 this identification can be purely coincidental, and more analysis is required to establish these results.

To summarize, in this section we presented the numerical calculations for the response function of generalized holographic superconductors
in 4D AdS-Schwarzschild black hole background, including effects of backreaction, and found negative refraction for low frequency.
In the next section, we will study response function in holographic superconductors in 4D single R-charged black hole background with backreaction.

\section{4D single R-charged black hole backgrounds}
\begin{figure}[t!]
\begin{minipage}[b]{0.5\linewidth}
\centering
\includegraphics[width=2.5in,height=2.0in]{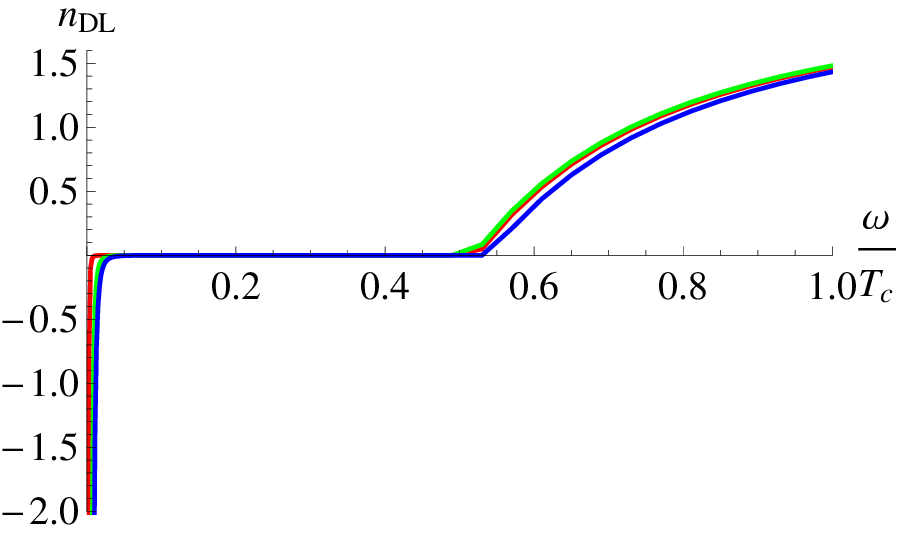}
\caption{$n_{DL}$ as a function of $\omega/T_c$ for different values of $\kappa$ with fixed $\xi=0$ at $T=0.5 T_c$. Here the red, green and blue curves correspond to $\kappa=$ $0.1$, $0.3$ and $0.5$ respectively.}
\label{RchargedNDLVsKappaZeta0TPt5Tc}
\end{minipage}
\hspace{0.4cm}
\begin{minipage}[b]{0.5\linewidth}
\centering
\includegraphics[width=2.5in,height=2.0in]{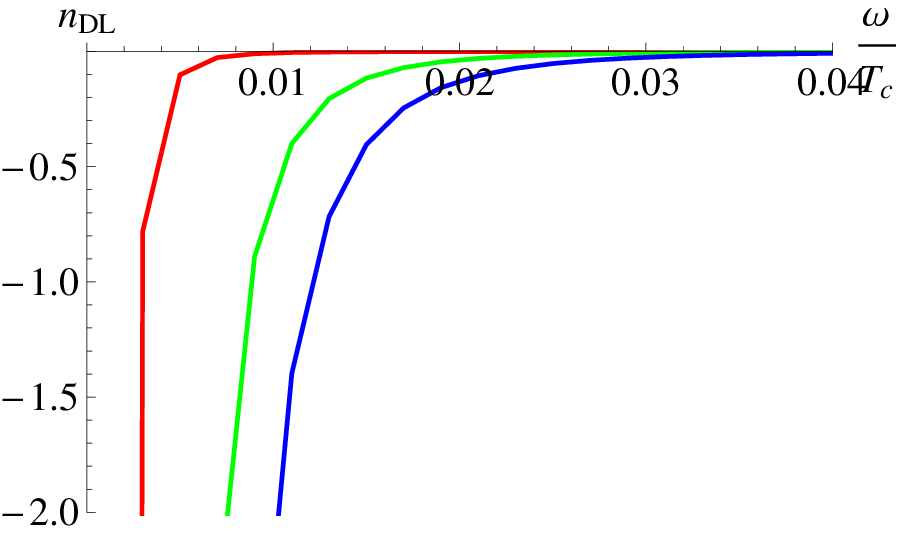}
\caption{Low frequency behavior of $n_{DL}$ as a function of $\omega/T_c$ for different values of $\kappa$ with fixed $\xi=0$ at $T=0.5 T_c$. Here the red, green and blue curves correspond to $\kappa=$ $0.1$, $0.3$ and $0.5$ respectively.}
\label{RchargedNDLsmallVsKappaZeta0TPt5Tc}
\end{minipage}
\end{figure}
\begin{figure}[t!]
\begin{minipage}[b]{0.5\linewidth}
\centering
\includegraphics[width=2.5in,height=2.0in]{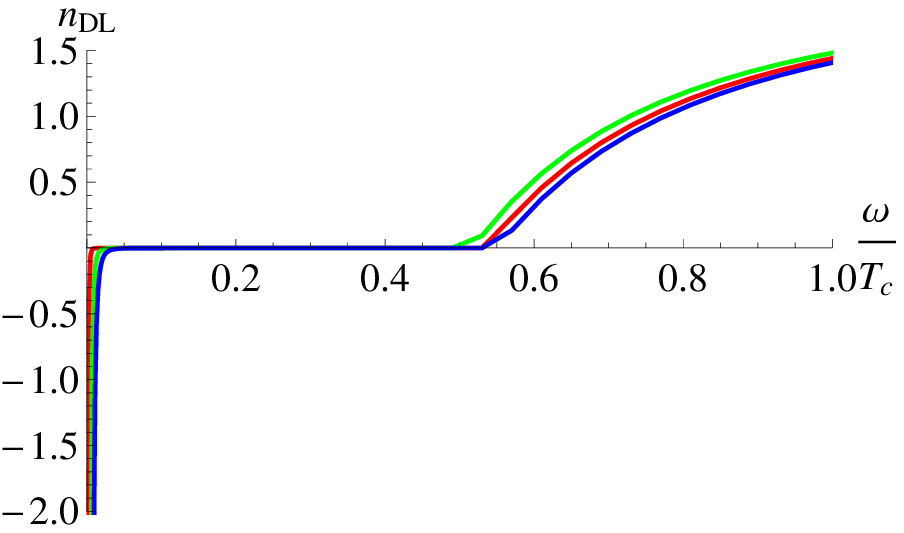}
\caption{$n_{DL}$ as a function of $\omega/T_c$ for different values of $\kappa$ with fixed $\xi=0.2$ at $T=0.5 T_c$. Here the red, green and blue curves correspond to $\kappa=$ $0.1$, $0.3$ and $0.5$ respectively.}
\label{RchargedNDLVsKappaZetaPt2TPt5Tc}
\end{minipage}
\hspace{0.4cm}
\begin{minipage}[b]{0.5\linewidth}
\centering
\includegraphics[width=2.5in,height=2.0in]{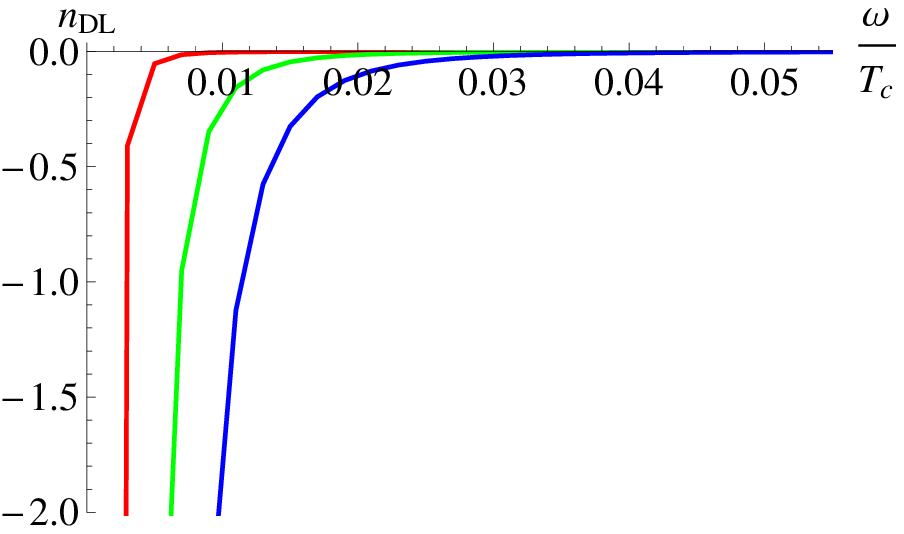}
\caption{Low frequency behavior of $n_{DL}$ as a function of $\omega/T_c$ for different values of $\kappa$ with fixed $\xi=0.2$ at $T=0.5 T_c$. Here the red, green and blue curves correspond to $\kappa=$ $0.1$, $0.3$ and $0.5$ respectively.}
\label{RchargedNDLsmallVsKappaZetaPt2TPt5Tc}
\end{minipage}
\end{figure}

For 4D single R-charged black hole backgrounds, the procedure for calculating response functions in generalized holographic superconductors is
entirely similar to what has been discussed in the previous section. The details of the computation have been relegated to Appendix A, and we simply
present numerical results here.

\begin{figure}[t!]
\begin{minipage}[b]{0.5\linewidth}
\centering
\includegraphics[width=2.5in,height=2.0in]{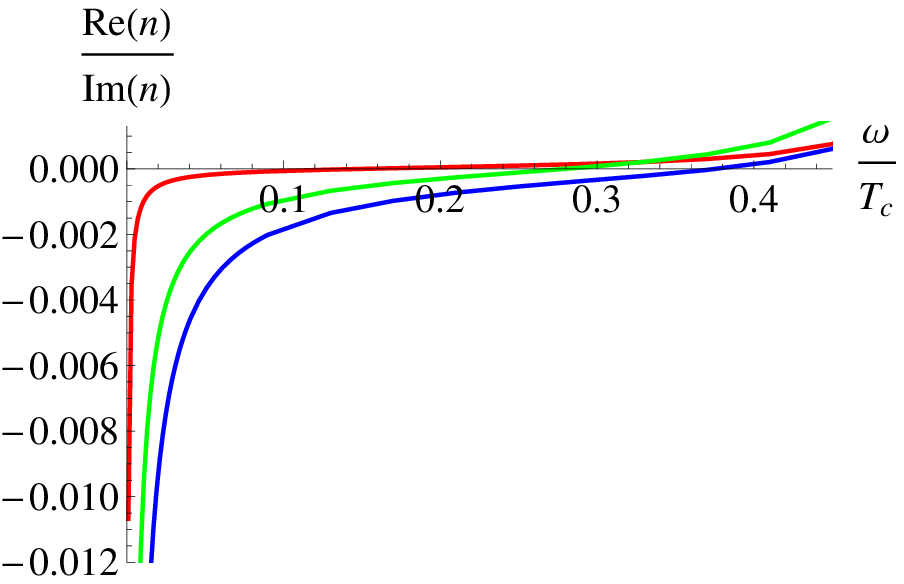}
\caption{Propagation to dissipation ratio as a function of $\omega/T_c$ for different values of $\kappa$ with fixed $\xi=0$ at $T=0.5 T_c$. Here the red, green and blue curves correspond to $\kappa=$ $0.1$, $0.3$ and $0.5$ respectively.}
\label{RchargedRenBYImnVsKappaZeta0TPt5Tc}
\end{minipage}
\hspace{0.4cm}
\begin{minipage}[b]{0.5\linewidth}
\centering
\includegraphics[width=2.5in,height=2.0in]{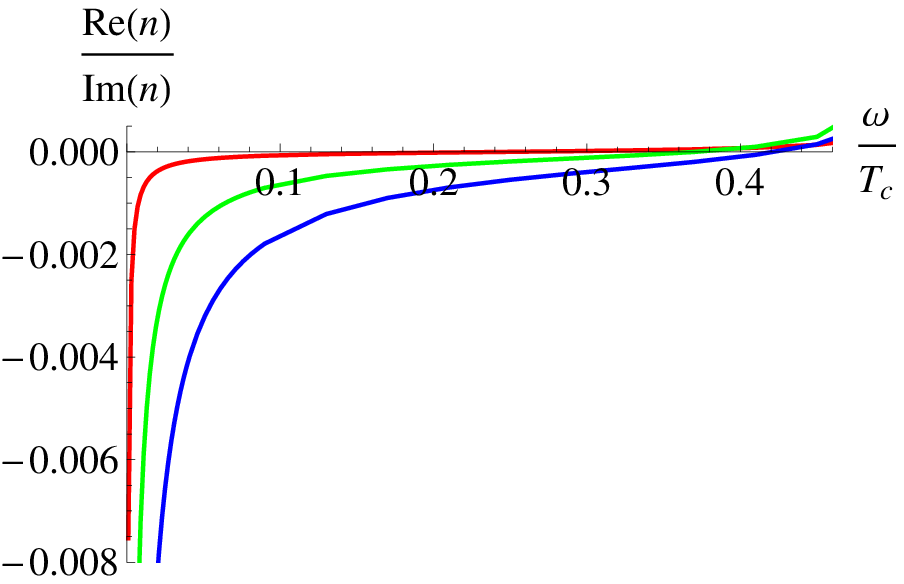}
\caption{Propagation to dissipation ratio as a function of $\omega/T_c$ for different values of $\kappa$ with fixed $\xi=0.2$ at $T=0.5 T_c$. Here the red, green and blue curves correspond to $\kappa=$ $0.1$, $0.3$ and $0.5$ respectively.}
\label{RchargedRenBYImnVsKappaZetaPt2TPt5Tc}
\end{minipage}
\end{figure}

The results for condensate, $\varepsilon$ and $\mu$ are qualitatively similar to case of AdS-Schwarzschild black hole and are therefore not presented.
The nature of $n_{DL}$ is shown in figs.(\ref{RchargedNDLVsKappaZeta0TPt5Tc}) and (\ref{RchargedNDLVsKappaZetaPt2TPt5Tc}) for two different
values of $\xi$.  Here again we choose $T=0.5T_c$ with the red, green and blue curves corresponding to $\kappa=$ $0.1$, $0.3$ and $0.5$ respectively.
We see that for both cases $n_{DL}<0$ at low frequency and once again we find negative refraction in the superconducting phase. This can be
clearly seen from figs.(\ref{RchargedNDLsmallVsKappaZeta0TPt5Tc}) and (\ref{RchargedNDLsmallVsKappaZetaPt2TPt5Tc}) where
$n_{DL}$ for low values of $\omega$ is shown. We find that the overall behavior of $n_{DL}$ in the superconducting phase in the R-charged background is
identical to that in the AdS-Schwarzschild background. However an important difference exists. The cutoff frequency where $n_{DL}$ goes from positive to
negative value, is now almost independent of $\kappa$. This can be observed by comparing the figs.(\ref{NDLVsKappaZeta0TPt5Tc}) and
(\ref{RchargedNDLVsKappaZeta0TPt5Tc}). We have also shown, in figs.(\ref{RchargedRenBYImnVsKappaZeta0TPt5Tc}) and
(\ref{RchargedRenBYImnVsKappaZetaPt2TPt5Tc}), the propagation to the dissipation ratio in the R-charged background. Here again we find large
dissipation in the frequency range where $n_{DL}$ is negative.

\section{Conclusions}
In this paper we have extended our study of response functions in generalized holographic superconductors in $(2+1)$ dimensions,
by considering the backreaction of the matter fields. We considered two distinct  backgrounds, namely 4D AdS-Schwarzschild and the 4D single R-charged black holes.
We first established the superconducting nature of the boundary theory and then studied its response functions under the external electromagnetic field and metric
perturbations. We applied the momentum dependent vector type perturbation and numerically found that, at low enough frequency, including effects of
backreaction makes $n_{DL}$ negative in both backgrounds. Our results also strengthen the general claim made in \cite{Amaritihydro} that
hydrodynamic systems which have gravity duals normally exhibit negative refraction below certain cutoff frequency. We further established the dependence
of the response functions on backreaction parameter $\kappa$ and model parameter $\xi$. We also found that most of the problems in the response
functions with probe limit are cured by backreaction effects. In particular, we showed that backreaction makes $Im(\mu)$ positive.
We further calculated the propagation to the dissipation ratio in our setup and found, as generally occur in metamaterials, small propagation.
However, higher backreaction seems to suggest enhance the propagation. We also presented a comparative analysis of the response functions in
the normal, superconducting and metastable regions of the phase diagram for our holographic superconductors. Our results provide predictions that
might be testable in realistic systems in future.

We conclude here by pointing out some problems which would be interesting to analyze in the future. It would be interesting to investigate the
response functions in holographic superconductors with backreaction by taking higher derivative interaction between the scalar and the gauge field into
account\cite{Dey1}. Even in the probe limit, the higher derivative interaction seems to provide unusual results for response functions\cite{Dey}.
Therefore it would certainly be interesting to analyse response functions with backreaction in this setup. Another interesting direction might be to investigate
response functions in striped holographic superconductors by introducing inhomogeneity in the system\cite{Flauger}. It may also lead us closer to
more realistic systems. We leave these issues for a future publication.

\begin{center}
{\bf Acknowledgements}
\end{center}
I sincerely thanks A. Amariti and D. Forcella for generously sharing their Mathematica code for optical properties of
$3+1$ dimensional holographic superconductors. The analysis in this paper is performed using a Mathematica routine that is inspired from this,
and is available upon request. I thank T. Sarkar for discussions and for a careful reading of the manuscript. I thank P. Phukon for useful discussions. Finally, I would like to thank A. Day and P. Roy for going through the
preliminary version of the manuscript and pointing out the necessary corrections. This work is supported by grant no. 09/092(0792)-2011-EMR-1 from CSIR, India.

\appendix
\section{Details of 4-D R-charged black hole backgrounds}

Here, we present the details of our calculations for response functions in 4D R-charged black hole backgrounds. We start with the action

\begin{eqnarray}
\textit{S}= \int \mathrm{d^{4}}x\! \sqrt{-g}\biggl[\frac{1}{2\kappa^{2}}\biggl(R+\frac{3}{L^{2}}\biggl(H^{1/2}+H^{-1/2}\biggr)\biggr)
-\frac{L^2 H^{3/2}}{8}\textit{F}_{\mu\nu}\textit{F}^{\mu\nu} &\nonumber \\ -\frac{3}{8}\frac{(\partial H)^2}{H^2} -\frac{1}{2}|\textit{D}\tilde{\Psi}|^{2} -\frac{1}{2}m^{2}|\tilde{\Psi}|^{2}
\biggl] \,
\label{actionRcharged1}
\end{eqnarray}
Writing the charged scalar field as $\tilde{\Psi}$ as $\tilde{\Psi}= \Psi e^{i \alpha}$ and following section $3$, we get the generalized action
\begin{eqnarray}
\textit{S}= \int \mathrm{d^{4}}x\! \sqrt{-g}\biggl[\frac{1}{2\kappa^{2}}\biggl(R+\frac{3}{L^{2}}\biggl(H^{1/2}+H^{-1/2}\biggr)\biggr)-\frac{L^2 H^{3/2}}{8}\textit{F}_{\mu\nu}\textit{F}^{\mu\nu} &\nonumber \\ -\frac{3}{8}\frac{(\partial H)^2}{H^2}  -\frac{(\partial_{\mu}\Psi)^2}{2} - \frac{m^2\Psi^2}{2}-\frac{|\textrm{G}(\Psi)|(\partial\alpha-q A)^2}{2}
\biggr] \,
\label{actionRcharged2}
\end{eqnarray}
taking the same ansatz for metric, scalar and gauge fields as in eq.(\ref{matric}) and (\ref{fieldsansatz}), we get the following equations of motion

\begin{eqnarray}
\Psi''+ \Psi '\left(\frac{g'}{g}- \frac{2}{z}+ \chi'\right) + \frac{ \Phi ^2 e^{-2\chi }}{2
   g^2}\frac{d\textrm{G}(\Psi)}{d\Psi} -\frac{m^2 \Psi }{z^2 g} =0
\label{scalareomRcharged}
\end{eqnarray}

\begin{eqnarray}
\Phi''- \Phi'\biggl(\chi'-\frac{3H'}{2H}\biggr) -\frac{2\textrm{G}(\Psi)}{z^2 g H^{3/2}}\Phi =0
\label{gaugeeomRcharged}
\end{eqnarray}

\begin{eqnarray}
g' - \kappa ^2 \left(\frac{z e^{-2 \chi} \textrm{G}(\Psi) \Phi^2}{2 g}+\frac{1}{2} z g
   \Psi'^2+\frac{m^2 \Psi^2}{2 z}+\frac{1}{4} z^3 e^{-2 \chi} H^{3/2} \Phi'^2 + \frac{3zgH'^2}{8H^2}\right) &\nonumber \\ -\frac{3g}{z}+\frac{3}{2z}(H^{1/2}+H^{-1/2})=0
\label{tteinsteineomRcharged}
\end{eqnarray}

\begin{eqnarray}
 \chi' + \frac{1}{2}z \kappa^2 \Psi'^2 + \frac{\kappa^{2}z e^{-2\chi} \Phi^2\textrm{G}(\Psi)}{2 g^2}+\frac{3z \kappa^2 H'^2}{8 H^2}=0
\label{rreinsteineomRcharged}
\end{eqnarray}

\begin{eqnarray}
H''+ H'\left(\frac{g'}{g}- \frac{2}{z}+ \chi'-\frac{H'}{H}\right) + \frac{e^{-2\chi}z^2H^{5/2}\Phi'^2 }{2g}+ \frac{H^{1/2}}{z^2\kappa^2g}(H-1) =0
\label{HeomRcharged}
\end{eqnarray}
Prime denotes a derivative with respect to $z$. We use the following boundary conditions in order to solve these five coupled differential equations. At the horizon
\begin{equation}
\Phi(1)=0 ,\ \ \Psi'(1)=\frac{m^2\Psi(1)}{g'(1)}.
\label{horizonbehavior4D}
\end{equation}
and at the boundary these fields asymptote to the following expressions
\begin{eqnarray}
\Phi=\mu-\rho z +... , ~~ \Psi=\Psi_{-}z^{\lambda_{-}}+\Psi_{+}z^{\lambda_{+}} + ... ~~ \chi\rightarrow 0, \ \ \ g\rightarrow 1+..., \ \ H\rightarrow 1+...
\label{boundarbehavior4D}
\end{eqnarray}
$\lambda_{\pm}=\frac{4\pm\sqrt{16+4m^2}}{2}$. Similar to the AdS-Schwarzschild black hole background case, here again we take $m^2=-2$ and treat $\Psi_{+}$ as the scalar operator and $\Psi_{-}$ as the source at the boundary.

For response function, we again consider the vector type perturbation $A_x(z)\neq 0$, $g^x_t(z)\neq 0$, $g^x_t(z)\neq 0$ with all other perturbation fields set to zero. From Maxwell equation we get
\begin{eqnarray}
A_{x}''+A_{x}'\biggl(\frac{g'}{g}+\chi'+\frac{3 H'}{2 H} \biggr)+A_{x}\biggl(\frac{\omega^2 e^{-2\chi}}{g^2} -\frac{k^2}{g}-\frac{2\textrm{G}(\Psi)}{z^2 g H^{3/2}}\biggr)+\frac{e^{-2 \chi}\Phi' g{^{x}_t}' }{g}=0
\label{Axeom4D}
\end{eqnarray}
and the $(t,x)$, $(x,y)$ and $(r,x)$-components of Einstein equation give
\begin{eqnarray}
g{^x_t}'' - g{^x_t}'\biggl(\chi'+ \frac{2}{z}\biggr)+ \kappa ^2 z^2 H^{3/2} A_x'\Phi'+ \frac{2 \kappa^2 \Phi \textrm{G}(\Psi) A_x}{g} - \frac{k^2 g{^x_t}}{g} - \frac{k \omega g{^x_y}}{g}=0
\label{gxteom4D}
\end{eqnarray}
\begin{eqnarray}
g{^x_y}''+ g{^x_y}'\biggl(\frac{g'}{g}+\chi' -\frac{2}{z}\biggr) +\frac{\omega^2 e^{-2 \chi} g{^x_y}}{g^2} +\frac{k \omega e^{-2 \chi} g{^x_t}}{g^2}=0
\label{gxyeom4D}
\end{eqnarray}
\begin{eqnarray}
\kappa^2 \omega z^2 H^{3/2} A_x \Phi'+ k g e^{2 \chi} g{^x_y}' +\omega
   g{^x_t}'=0
\label{constraineom4D}
\end{eqnarray}
With eqs.(\ref{Axeom4D})-(\ref{constraineom4D}) in hand, we calculated current-current correlators and hence all the response functions of holographic superconductors in R-charged black hole background by implementing the analogous procedure mentioned in section $4$.

\end{document}